\documentclass[manuscript,natbib]{aastex}

\usepackage{subfigure}  

\newcommand{\oi}{[O~{\sc i}]}
\newcommand{\neii}{[Ne~{\sc ii}]}
\newcommand{\uflux}{erg\,cm$^{-2}$\,s$^{-1}$}  

\slugcomment{$^\star$Based on observations made with VISIR on the UT3/Melipal ESO Telescope at Paranal under programme ID 084.C-0088(A)}

\shorttitle{}
\shortauthors{}

\begin{document}
\bibliographystyle{apj}

\title{The Photoevaporative Wind from the Disk of TW~Hya}


\author{I. Pascucci\altaffilmark{1}}
\affil{Lunar and Planetary Laboratory, The University of Arizona, Tucson, AZ 85721, USA}
\email{pascucci@lpl.arizona.edu}
\altaffiltext{1}{Department of Physics and Astronomy, Johns Hopkins University, Baltimore, MD 21218}

\author{M. Sterzik}
\affil{European Southern Observatory, Casilla 19001, Santiago 19, Chile}

\author{R. D. Alexander}
\affil{Department of Physics \& Astronomy, University of Leicester, University Road, Leicester, LE1 7RH, UK}

\author{S. H. P. Alencar}
\affil{Departamento de Fisica - ICEx  – UFMG, MG, Brazil}

\author{U. Gorti\altaffilmark{2}}
\affil{SETI Institute, 189 Bernardo Ave., Mountain View, CA 94043, USA}
\altaffiltext{2}{NASA Ames Research Center, Moffett Field, CA 94035, USA}

\author{D. Hollenbach}
\affil{SETI Institute, 189 Bernardo Ave., Mountain View, CA 94043, USA}

\author{J. Owen}
\affil{Institute of Astronomy, Madingley Road, Cambridge CB3 0HA}

\author{B. Ercolano}
\affil{University Observatory Munich, D-81679, Munich, Germany}

\author{S. Edwards}
\affil{Astronomy Department, Smith College, Northampton, MA 01063 USA}




\begin{abstract}
Photoevaporation driven by the central star is expected to be a ubiquitous and important mechanism to disperse the circumstellar dust and gas from which planets form. Here, we present a detailed study of the circumstellar disk surrounding the nearby star TW~Hya and provide observational constraints to its photoevaporative wind. Our new high-resolution ($R\sim30,000$) mid-infrared spectroscopy in the \neii{} 12.81\,\micron{} line confirms 
that this gas diagnostic traces the unbound wind component within 10\,AU from the star. From the blueshift and asymmetry in the line profile, we estimate that most ($>$80\%) of the \neii{} emission arises from disk radii where the midplane is optically thick to the redshifted outflowing gas, meaning beyond the 1 or 4\,AU dust rim inferred from other observations. We re-analyze high-resolution ($R\sim48,000$) archival optical spectra searching for additional transitions that may trace the photoevaporative flow. Unlike the \neii{} line, optical forbidden lines from OI, SII, and MgI are centered at the stellar velocity and have symmetric profiles. The only way these lines could trace the photoevaporative flow is if they arise from a disk region physically distinct from that traced by the \neii{} line, specifically from within the optically thin dust gap. However, the small ($\sim$10\,km/s) FWHM of these lines suggest that most of the emitting gas traced at optical wavelengths is bound to the system rather than unbound. We discuss the implications of our results for a planet-induced versus a photoevaporation-induced gap.
\end{abstract}





\section{Introduction}

One of the yet unsolved questions in planet formation is how young stars lose their disks. Planet formation certainly contributes to the clearing of primordial disk material. Its first steps, the coagulation of sub-micron sized grains and their subsequent settling to the disk midplane, have been identified toward many
protoplanetary disks (e.g., \citealt{natta07} for a review). 
However, planet formation is likely not the major disk dispersal mechanism, as evinced by the small total mass in the planets (both in our solar system and in other planetary systems) when compared to the mass of protoplanetary disks (e.g., \citealt{hayashi85}). Furthermore, a disk dispersal mechanism based only on planet formation would require shorter dispersal timescales in high metallicity environments \citep{ercolanoclarke10}, which is not supported  by recent observations of disk fractions in low metallicity clusters of the extreme outer Galaxy \citep{yasui09}.

Current models of protoplanetary disk evolution suggest that viscous evolution (accretion of gas onto the central star) and photoevaporation driven by the central star (heating of disk gas to thermal escape velocities) are the main disk dispersal mechanisms (e.g., \citealt{hollenbach00,gortietal09,owen10}). Indeed, there is abundant observational evidence that young stars are accreting disk gas: from the optical and ultraviolet excess continuum emission arising from the accretion shock (e.g., \citealt{calvet00}) to the broad permitted emission lines from infalling gas (e.g., \citealt{muzerolle98}). Measurements of mass accretion rates using these diagnostics are now available for hundreds of stars in nearby star-forming regions and qualitatively follow the time evolution predicted by these models \citep{hartmann98,sicilia10}. However, at least two other observables are not reproduced by models of viscously evolving disks. 
First, the disk dispersal timescale, which is measured to be of just a few Myr, is predicted to be too long using the viscosity inferred from mass accretion rate measurements. Both optically thick dust disks as well as gaseous disks should be numerous and detectable around stars that are $\sim$10\,Myr-old, which contrasts with  observations (e.g., \citealt{pascucci10}). Second, the transition from disk-bearing to disk-less stars is observed to be quick (just $\sim$10$^5$ years), too quick to be explained by viscous evolution alone (see e.g., \citealt{armitage10} for a review).

These observables led theorists to propose photoevaporation driven by the central star as the next ubiquitous and most relevant mechanism to disperse protoplanetary disks (e.g., \citealt{hollenbach00,dullemond07}). In the first model combining the effect of accretion and photoevaporation \citep{clarke01}, extreme UV photons (EUV, 13.6\,eV\,$< h \nu \le 100$\,eV) from the central star heat and ionize the disk surface. Beyond the radius at which  the sound speed of the gas equals the local Keplerian orbital speed (gravitational radius, $r_{\rm g}$), 
the hot gas becomes  unbound and  a photoevaporative wind is established. When the disk accretion rate through $r_{\rm g}$ drops below the wind loss rate, photoevaporation limits the supply of gas to the inner disk, which drains onto the star on the
local viscous timescale  -- of order 10$^5$ years. Many theoretical developments have been made since this first model. First,  \cite{liffman03}, \cite{font04}, and \cite{adams04} have included the effects of angular momentum support and showed that  photoevaporation occurs within $r_{\rm g}$ and  happens mostly outside $\sim$0.15\,$r_{\rm g}$ (hereafter critical radius). Second, \cite{alexander06a,alexander06b} showed that the direct EUV irradiation of the disk rim can drastically reduce the disk lifetime once the inner disk is cleared out. Third, \cite{ercolano08}, and \cite{gorti09} have recognized that stellar Xrays and FUV photons can launch winds from denser disk regions and thus dramatically increase the wind mass-loss rate produced by EUV photons alone.

Recently, several papers have discussed whether other disk properties, beyond disk lifetimes, are consistent with photoevaporation models. There is yet no consensus: \citet{kim09} include photoevaporation as a possible explanation for some of the transition disks in their sample; \citet{currie09} argue that the large number of evolved dust disks at 5\,Myr is inconsistent with photoevaporation; \citet{cieza08} find support for photoevaporation based on the lack of optically thin inner disks with detectable dust outer disks. All these studies and similar ones relied on comparing model predictions with the properties of dust disks. A more direct way to test photoevaporation is to observe the gas disk component, especially that component that is not gravitationally bound and is part of the photoevaporating flow. 

Following this approach, \citet{pascucci09} have shown that the \neii{} line at 12.81\,\micron{} 
is a good diagnostic for the photoevaporative wind in some systems that have transition-like spectral energy distributions (SEDs). Such systems display small near-infrared excess emission but large mid- to far-infrared emission and are thus thought to be in the process of clearing out their primordial disk material. 
The most convincing case for photoevaporation is that for the almost face-on transition disk around the
10\,Myr-old star TW~Hya. The measured blueshift of $\sim$6\,km/s demonstrates that there is unbound gas
leaving the disk and moving toward the observer \citep{pascucci09}. If the wind is fully ionized, the measured \neii{} flux translates into a mass-loss rate of $\sim 10^{-10}$M$_\sun$/yr.  However, if the wind is mostly neutral, as predicted by models of Xray photoevaporation\footnote{Gorti \& Hollenbach (2009) point out that Xrays alone cannot drive a strong wind but it is rather the combination of FUV photons (which dissociate the molecules and thereby reduce the ability of the disk gas to cool) and Xrays (which heat the gas) that drive the wind}, then the \neii{} emission is a trace component of the wind and the observed line strength and profile imply a much larger wind rate up to $\sim 10^{-8}$M$_\sun$/yr (\citealt{eo10}). We also note that this relatively old star is still accreting disk gas at a rate that is about an oder of magnitude lower than that of classical T Tauri stars ($\sim$10$^{-9}$\,M$_\sun$/yr, Muzerolle et al. 2000 and  Alencar \& Batalha 2002).

Being at only 51\,pc \citep{mamajek05} and clearly presenting evidence for photoevaporation,  TW~Hya is the ideal target to provide additional observational constraints to models of disk photoevaporation and to help determine wind mass-loss rates. In this paper we present new observations of the \neii{} 12.81\,\micron{} line toward TW~Hya and employ the spectroastrometry technique to estimate the radii traced by the ionized wind component. We also analyze archival optical spectra to search for the neutral disk wind component predicted by Xray-driven photoevaporation models. We discuss the implications of our results on the gas distribution in the inner disk of TW~Hya and speculate on how that structure might link to planet formation and photoevaporation.


\section{Photoevaporative disk wind models}\label{Sect:PhotModels}
Before presenting our observations and analyzing the infrared and optical spectra, we wish to briefly summarize the main features of photoevaporative disk wind models and point out their major differences. Our goal is to clarify what observables can be used to characterize the structure and extension of the wind based on model predictions.

There are basically three main sources of ionization and heating of the disk surface: Xrays (0.1\,keV\,$< h \nu <$\,10\,keV), EUV (13.6\,eV\,$< h \nu <$\,0.1\,keV), and FUV (6 eV $< h \nu <$ 13.6 eV) photons.  
The first and simplest  models assumed that photoevaporation is driven solely by stellar EUV photons. These photons ionize the hydrogen in the very upper layers of the disk and heat it to a temperature of $\sim$10$^4$\,K, independent of radius, thus creating a kind of coronal H~II region (Hollenbach et al. 1994, Alexander et al. 2006a,b). 
This fully ionized layer will produce forbidden lines at infrared and optical wavelengths whose relative intensities depend on the intensity and the spectral energy distribution of the EUV field impinging on the disk (Hollenbach \& Gorti 2009), which is not constrained observationally. Assuming plausible stellar ionizing fluxes of $\sim 10^{41}$\,photons/s and a solar abundance of the elements, EUV-only models predict a relatively strong \neii{} emission line at 12.81\,\micron{} (Alexander 2008, Hollenbach \& Gorti 2009) and similarly strong [S~II] lines at 6731 and 6716\,\AA{} and [N~II] lines at 6583\,\AA{}  (Font et al. 2004). Because the fraction of neutral gas is very low in the EUV layer, forbidden lines from atomic species, such as the [O~I] line at 6300\,\AA{}, are expected to be weak (Font et al. 2004, Hollenbach \& Gorti 2009). The most recent developments of the EUV-only models include  a detailed treatment of the flow structure and predict line profiles from the wind in addition to line intensities (Alexander 2008).  As we will show in Sect.~\ref{Sect:modelwind}, these simple models can also produce channel maps at specific emission lines that can be used to constrain the spatial distribution of the wind.



More comprehensive models, such as those developed by Ercolano et al. and Gorti \& Hollenbach, include other sources of disk heating and ionization in addition to stellar EUV photons. Specifically, the most recent development of the Ercolano et al. models includes Xray and EUV heating and ionization of the atomic disk gas and a 2D hydrodynamic calculation wich allows predicting line intensities and profiles from a variety of forbidden and semi-permitted transitions in the wind (Ercolano \& Owen 2010). Hereafter, we will call these models X-EUV models. Gorti \& Hollenbach~(2009) include all three main heating and ionization sources and a detailed treatment of molecular cooling in their thermal balance calculations. For this more complex model, which we will call XE-FUV model, hydrodynamical calculations have not been perfomed yet hence we will limit the comparison between model and observations to line intensities. 
The X-EUV and XE-FUV models further differ in their assumptions of the stellar Xray and EUV spectra and therefore also in the predicted mass-flow rates. Regardless of these differences, the inclusion of Xray and FUV heating results in a predominantly denser, but cooler (hence mainly neutral), photoevaporative wind compared to the EUV-only models and the mass loss rates are higher by one (XE-FUV) or two (X-EUV) orders of magnitude. Depending on the star/disk parameters, the [OI] emission line at 6300\,\AA{}, whose low-velocity component is always detected toward accreting T Tauri stars (Hartigan et al. 1995), may trace the neutral component of the disk wind (Ercolano \& Owen 2010).  Other forbidden lines that are expected to be strong in the Xray ionized wind are the  \neii{} line at 12.81\,\micron{} (Hollenbach \& Gorti 2009, Ercolano \& Owen 2010) and the  [S~II] lines at 6731 and 6716\,\AA{} (Ercolano \& Owen 2010). 

We will now analyze our mid-infrared and optical spectra focusing on gas lines that could trace the ionized and neutral components of the disk wind.

\section{New mid-infrared observations and data reduction}\label{Sect:Observations}
Our first VLT/VISIR observations have shown that  the \neii{} 12.81\micron{} emission from
TW~Hya traces unbound gas moving toward the observer, likely in a photoevaporative disk wind
\citep{pascucci09}.
We performed additional long-slit high-resolution spectroscopy  with VISIR  \citep{lagage04} on 23-24
February 2010 attempting to detect the photoevaporative flow at a resolution better than the instrument resolving power
($\sim$\,400\,mas or 20\,AU at the distance of TW~Hya). 
To reach this goal we have applied the so-called spectro-astrometric technique (e.g.,
\citealt{wg08}) and acquired spectra at four different slit orientations (position angle PA = 0, 270, 333,
63\degr)
 interleaved by four antiparallel orientations, i.e. where the slit is rotated by 180\degr{} (position angle PA = 180, 90, 153, 243\degr), see Table~\ref{table:log}.
 PAs 0\degr{} and 180\degr{} correspond to N-S slit orientations on the sky, 270\degr{} and 90\degr{} to W-E while the other two couples of PAs were chosen to cover the PA of the disk of TW~Hya (i.e. the angle between N and the major axis of the disk measured E of N, see Table~\ref{tab:twhyavalues}) and a direction perpendicular to it.
As in past observations, we used a slit width of 0\farcs4 which provides a resolution of R$\sim$30,000 or
$\sim$10\,km/s in velocity scale  at $\sim$12.8\micron{} \citep{kaufl06,pascucci09}.

\begin{figure}
\includegraphics[scale=0.8]{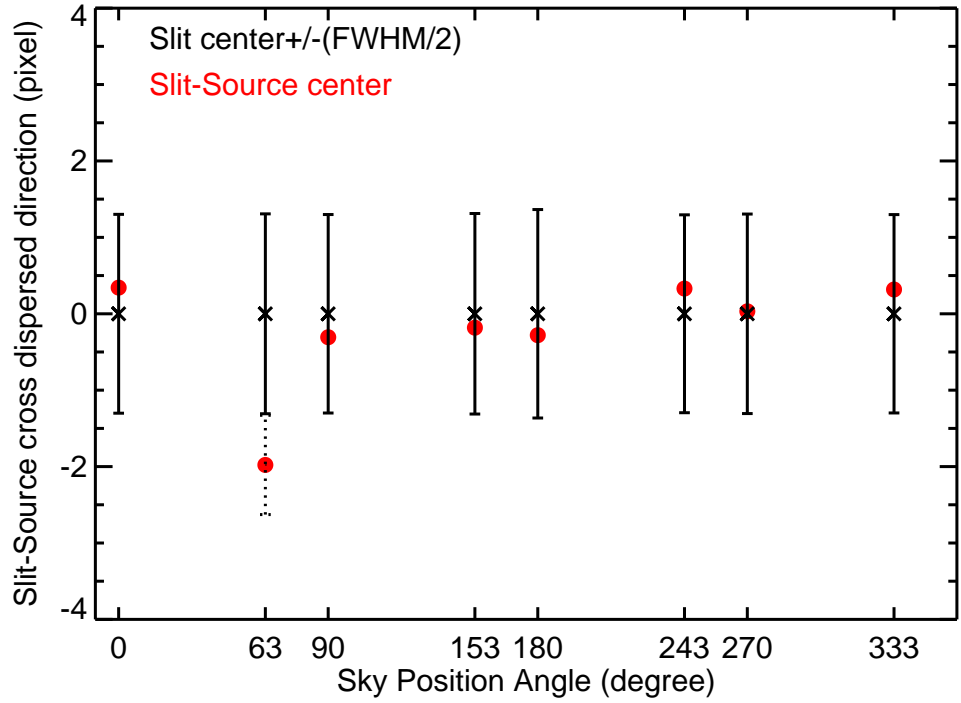}
\caption{Difference in pixels between the location of TW~Hya and the center of the slit before each spectroscopic integration. The x-axis gives the position angle in degrees. Vertical errorbars centered at y=0 show the extension of the slit image (FWHM) in cross-dispersion direction. Except for PA=63\degr{} TW~Hya was well positioned inside the slit. \label{fig:loc_source}
}
\end{figure}

Before each spectroscopic integration we acquired TW~Hya with the PAH2-NEII filter to accurately position the source in the slit.  Fig.~\ref{fig:loc_source} shows the difference in pixels between the location of the target and the slit center in cross-dispersion direction, i.e. perpendicular to the slit length. The FWHM of the slit ($\sim 3$\,pixels) is also overplotted. Except for PA=63\degr{}, TW~Hya was positioned within half of the slit FWHM from the slit center. We note, however, that for this specific PA the emission is also the faintest, hence the error in the source location is the largest (dashed line in Fig.~\ref{fig:loc_source}). The source faintness could be due to flux loss caused by the imperfect centering.

We applied the standard chopping/nodding technique with a throw of 8\arcsec{} along the slit  to suppress the mid-infrared background. Based on our previous experience, we set the on-source exposure time to 1\,h for each slit orientation. For all PAs in the first night and for the first PA in the second night we also observed a standard star immediately before or after the TW~Hya spectroscopic exposure and with the same slit PA to correct the spectra for telluric absorption and to obtain an absolute flux calibration. A summary of the observations is presented in Table~\ref{table:log}.

Each spectroscopic exposure has been reduced as described in \citet{pascucci09} using the VISIR pipeline
version 3.2.2 (Lundin 2008). In brief,  images are first corrected by fluctuations in the background using the off-source chop and nod exposures. In parallel, a reference frame of the infrared background is also created. Next,  images are corrected for the optical distortion, and are shifted and added to form a final combined image from which the spectrum is extracted following the optimal extraction method by \citet{horne86}. Finally, the spectrum is wavelength calibrated by cross correlating the background frame spectrum to a synthetic model spectrum of the atmosphere in the observed wavelength range (note that the resulting wavelengths are in vacuum).

\begin{figure}
\includegraphics[scale=0.4]{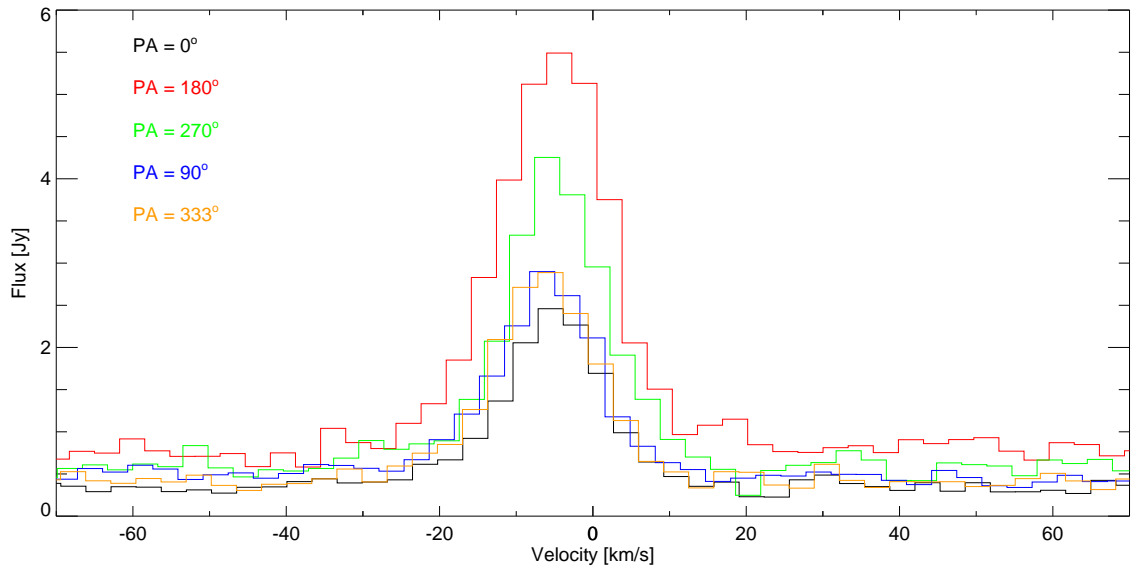}
\caption{The five flux calibrated spectra of TW~Hya around the \neii{}
emission line. The x-axis gives the velocity in the stellocentric frame.\label{fig:neiifluxes}
}
\end{figure}

To remove telluric features we have used ATRAN model atmospheres for different elevations and amounts of precipitable water as follows. We normalized each spectrum of TW~Hya to the continuum, such that the continuum is 1, and scaled the model atmospheres to minimize (in a $\chi^2$ statistics) the difference between the observations and the model in regions where there are strong atmospheric absorption features (the strongest one is from CO$_2$ around 12.812\,\micron). During the minimization we took into account the flux uncertainty at each wavelength and also allowed for a small shift of $\pm$1 pixel in wavelength to find the best scaled/shifted atmospheric model that matches the continuum of TW~Hya. Since in the first night we also acquired standard stars immediately before or after each PA of the TW~Hya spectra, we could verify that the approach described above gives identical results (within the flux uncertainties) to the more traditional approach of dividing each science spectrum by that of a flux calibrated standard star spectrum. Fig.~\ref{fig:neiifluxes} presents the five TW~Hya spectra that could be flux calibrated while Table~\ref{tab:var} summarizes the main properties of the observed \neii{} lines for all PAs. 

\subsection{Accuracy of the VISIR wavelength solution}\label{Sect:waveaccuracy}
\begin{figure}
\includegraphics[scale=0.4]{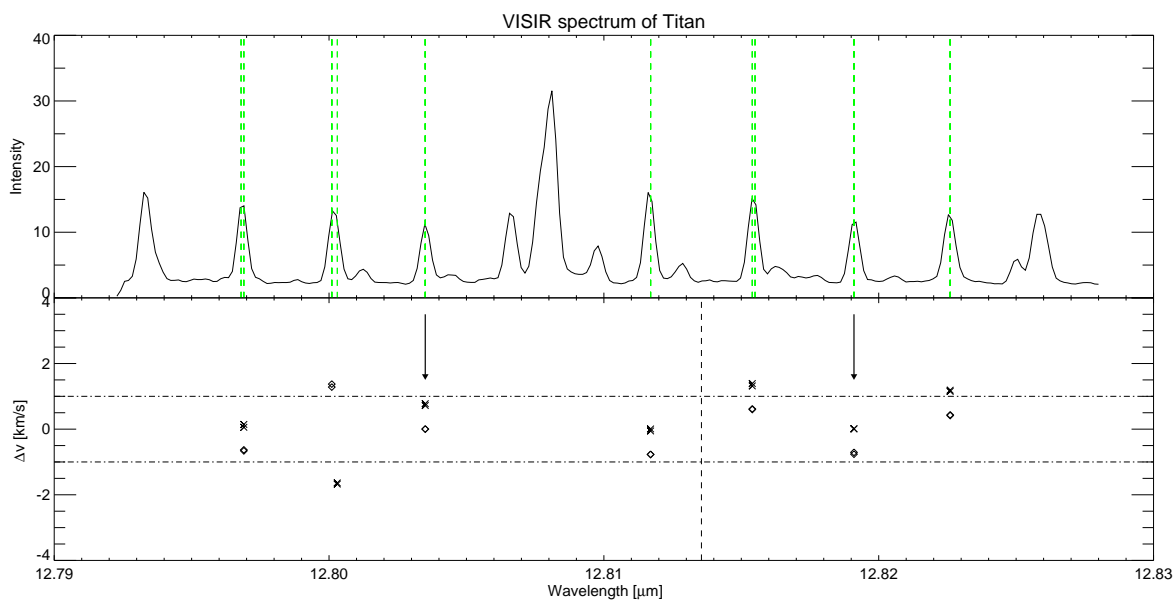}
\caption{VISIR spectra of Titan demonstrating the accuracy of the dispersion solution. The upper panel marks in (green) dashed-lines the C$_2$H$_2$ transitions from the HITRAN database. This spectrum of Titan (we analyzed two spectra) has been shifted to match the single  C$_2$H$_2$ line at $\sim$12.803\micron. The lower panel gives the velocity difference in km/s for lines marked in the upper panel. The location of the \neii{} line is also shown (black vertical dashed line). The two single C$_2$H$_2$ lines are marked with arrows.
'X' symbols give shifts when the rest frame is computed from the single line at $\sim$12.819\micron{}, while diamonds from the single line at $\sim$12.803\micron. The standard deviation of the velocity difference, which gives the relative uncertainty, is 1\,km/s. The absolute uncertainty is a factor of two larger (see text). \label{fig:titan_peaks}
}
\end{figure}

Because one of our goals is to measure small shifts in the peak of the \neii{} line, we performed additional tests to verify the instrument wavelength solution and investigate the relative and absolute wavelength accuracy of VISIR in the high-resolution long-slit mode. First of all, we reduced and analyzed 8 additional archival spectra from 3 different infrared-bright K4-6 III standard stars (HD~136422, HD~139127, HD~149447) observed in the same setting as TW~Hya.  After applying the proper heliocentric correction and shifting the spectra in the stellocentric frame, we compared the observed peak position of two photospheric absorption lines to those predicted by the MARCS model atmosphere of a K5 III star (HR 6705; the spectrum was kindly provided by L. Decin).
While differences in velocities are confined within $\pm$2\,km/s it is apparent that one of the lines is typically redshifted 
while the other 
is typically blueshifted. 
To investigate whether there could be an issue in the wavelength dispersion solution and whether errors could be larger at other wavelengths, we turned to spectra of Titan which were acquired with the same setting as TW~Hya (VLT/VISIR PID~083.C-0883). Titan spectra present many unresolved lines from C$_2$H$_2$ and C$_2$H$_6$ molecules in this setting. In case of large drifts in wavelength, we might see the FWHM of unresolved lines to change across the spectrum. We measured the FWHMs of 7 unresolved emission lines in two spectra of Titan covering the full VISIR wavelength range (3 before and 4 after the \neii{} line in TW~Hya) and found no such trend, rather we report a narrow range of FWHMs with a mean value of 10\,km/s, the spectral resolution of VISIR in this setting. In addition, we compared the location of Titan lines with the rest wavelengths reported in the HITRAN database \citep{rothman09}. In the setting covered with VISIR there are seven transitions from the C$_2$H$_6$ molecule, 2 of which are single and can be used to measure the relative centroids to other lines through the spectrum. Fig.~\ref{fig:titan_peaks} shows that the relative uncertainty between the lines is 1\,km/s, one-tenth of the spectral resolution of VISIR in this mode. However, as hinted by the test on the standard stars and as we will see on the TW~Hya data (Sect.~\ref{sect:results}), the absolute accuracy reachable with VISIR is likely a factor of 2 higher (2\,km/s), due to uncertainties in placing the source in the center of the slit.





\section{Constraints on the wind structure from ionized and neutral species}\label{sect:results}
In the following we analyze our new mid-infrared high-resolution spectra 
with special emphasis on characterzing the \neii{} peak velocities and the overall line profile. 
We also re-analyze already published high-resolution optical spectra of TW~Hya to search for additional transitions from neutral and ionized species that may trace the photoevaporative flow.
The optical spectra were obtained between 1998 and 2000 with the FEROS echelle spectrograph on the 1.5\,m ESO telescope on La Silla at a resolution of $\sim 6$\,km/s ($R\sim$48,000). \citet{alencar02} presented an extensive analysis of the optical variability of TW~Hya based on this dataset. We refer to their paper for details on the data reduction.

\subsection{\neii{} line fluxes and peak velocities}\label{Sect:var}
{\it Spitzer}/IRS spectra of TW~Hya taken in three different years show variations of up to $\sim$30\% in the continuum emission, as well as in the \neii{} flux and equivalent width \citep{najita10}. These differences may arise from pointing errors and/or true source variability. The VISIR spectra we acquired in this observational campaign provide additional epochs to further investigate continuum and line variability on the timescale of years as well as on the much shorter timescale of hours. 

For each VISIR spectrum we compute the location of \neii{} peak emission in the stellocentric frame (after assuming a stellar radial velocity of 12.2$\pm$0.5\,km/s, \citealt{weintraub00}), the EW, and when standard stars were acquired close in time to the TW~Hya spectroscopic exposures, also the line flux and continuum near the line (essentially for each PA in the first night and for the first PA in the second night). Peak positions and \neii{} line fluxes are calculated from gaussian fits to the observed line profiles, while EWs are computed by integrating the line flux within $\pm 4 \sigma$ of the fitted gaussian.
Table~\ref{tab:var} summarizes the main results and Fig.~\ref{fig:neiifluxes} shows line profiles for the five flux calibrated spectra.

\begin{figure}
\includegraphics[scale=0.8]{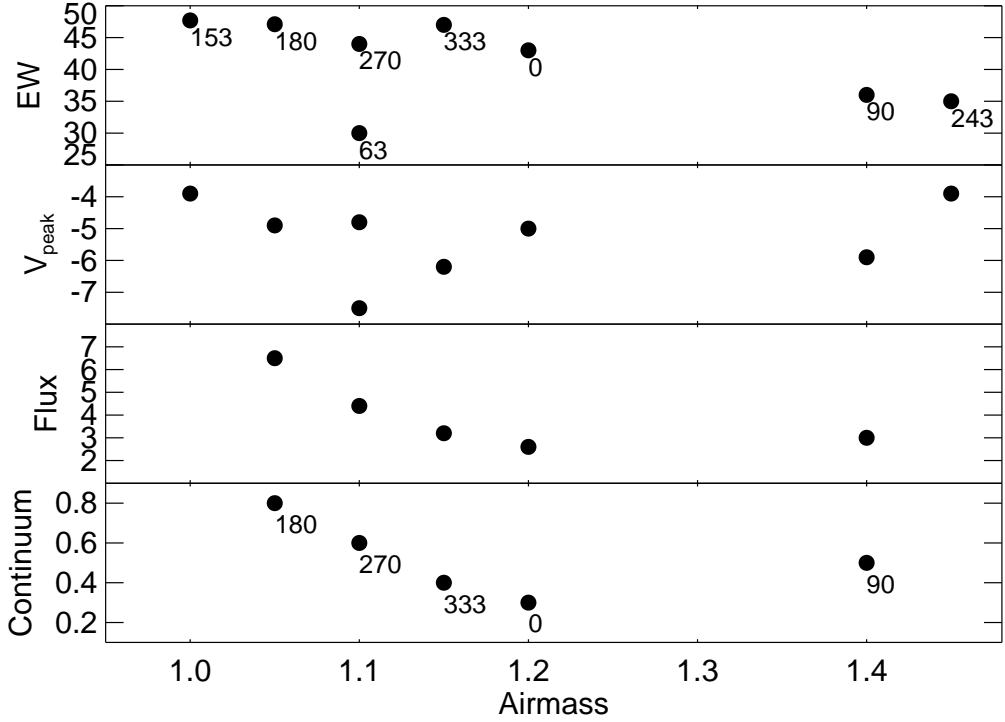}
\caption{\neii{} line EW (\AA), peak centroid (km/s in the stellocentric frame), flux ($\times 10^{-14}$\uflux), and adjacent continuum (Jy) versus the airmass at which the spectroscopic observations were carried out. PAs are also overplotted.
Note the anti-correlation between the \neii{} flux and  adjacent continuum with airmass suggesting that most of the observed variations in these quantities are related to flux calibration uncertainties rather than to intrinsic source variability. Also note that for PA=63\degr, which shows the lowest EW and the largest blueshift, the source was not well centered in the slit (see Fig.~\ref{fig:loc_source}). \label{fig:corr}
}
\end{figure}

To test the hypothesis of variations due to calibration uncertainties, we search for correlations between the main properties of the \neii{} line and the adjacent continuum with the airmass and the location of the source in the slit. We find an anti-correlation between the line flux and continuum and  the source airmass (lower airmass$\rightarrow$higher flux and continuum, see Fig.~\ref{fig:corr}) suggesting that most variations in these quantities are due to flux calibration uncertainties. We do not find any correlation between these quantities and the location of the source in the slit. However, we note that the displacement of the source with respect of the center of the slit might be the cause for the lowest EW and the largest blueshift in the \neii{} line for PA=63\degr{} (see Fig.~\ref{fig:loc_source} and Fig.~\ref{fig:corr}). Unfortunately, we do not have a standard star to flux calibrate the TW~Hya spectrum at this PA but our expectation is of a reduced line flux and continuum. We also note that the highest \neii{} flux we measure (for PA=180\degr, smallest airmass) is as high as that measured by {\it Spitzer}/IRS within the estimated errorbars. This demonstrates that the small VISIR slit recovers all the \neii{} emission from the disk of TW~Hya implying that the emission is confined within $\pm$10\,AU from the star.

\neii{} line EWs are less affected by differences in the airmass: for the 5 exposures at airmass
$< 1.3$ (excluding PA=63\degr{}) the mean EW is 46$\pm$2\AA{}. This value is
slightly larger than the \neii{} EW from our 2008 high-resolution spectrum
\citep{pascucci09}, the same as the 2006 EW measured with {\it Spitzer}/IRS
\citep{najita10}, and smaller by about 20\,\AA{} than the 2004 and 2008 EWs measured
with {\it Spitzer}/IRS  and the 2007 EW measured with Gemini/Michelle by \citet{herczeg07}. Although the sampling is sparse, the \neii{} EW essentially oscillates around two values ($\sim$45 and 65\,\AA{}) which suggests intrinsic variability in the \neii{}
emission or nearby continuum on a timescale of $\sim$1-2 years.

We also confirm the blueshift in the \neii{} line with respect to the radial velocity of TW~Hya (see
Fig.~\ref{fig:corr} second panel). The mean blueshift is -5$\pm$1\,km/s in agreement with our previously
reported blueshift \citep{pascucci09}. The peak position of the \neii{} line is not correlated with the airmass at which the source was observed, nor with the location of the source in the slit (but see previous note for PA=63\degr). 
The FWHM of the \neii{} line presents the least variations with PA, is not correlated with the EW, peak position, nor with the airmass: the mean value and standard deviation for the FWHM are 16.3$\pm$0.9\,km/s, or an intrinsic FWHM of 12.9$\pm$0.9\,km/s after deconvolution with the instrumental FWHM of VISIR of 10\,km/s.

\subsection{Forbidden and semi-permitted optical lines}\label{Sect:otherlines}

We search the FEROS optical spectra for all the lines predicted by \cite{font04}, \citet{hollenbach09}, and \citet{eo10}  to have
luminosities of a few 10$^{-7}$\,L$_\sun$ or higher, thus comprising forbidden and semi-permitted transitions from OI, OII, MgI, NII, and SII.
We report equivalent widths and upper limits in Table~\ref{tab:optlines} for representative transitions. To search for weak lines and properly
measure EWs we had to correct for the numerous photospheric absorption lines present in the optical spectra of TW~Hya.  We followed the
approach described by \citet{alencar02} and  divided the continuum normalized spectra of TW~Hya with the artificially veiled and continuum normalized spectrum of GJ~1172, a K7 V star with
similar {\it v sini} as TW~Hya (see also Hartigan et al. 1989 for the procedure and definition of veiling). In addition, we combine all observed spectra in a median high signal-to-noise spectrum normalized to the continuum
level.

\begin{figure}
\includegraphics[scale=0.8]{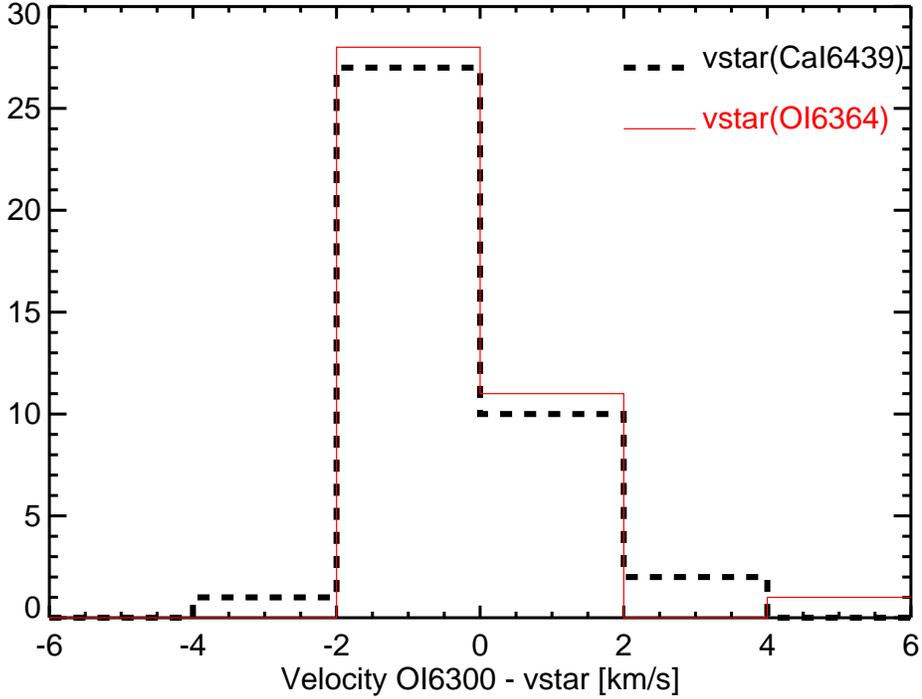}
\caption{Histogram of the velocity shifts of the \oi{} 6300\,\AA{} line with respect to the stellar velocity measured via the photospheric  CaI 6439 (black dashed line) and the \oi{} 6364\,\AA{} (red solid line) lines. The  CaI 6439\,\AA{} and the  \oi{} 6364\,\AA{} fall on the same echelle order while the  \oi{} 6300\,\AA{} line falls on a different echelle order. The fact that the two histograms look very similar and have the same median and standard deviation  demonstrate that the \oi{} 6300\,\AA{} line is centered at the stellar velocity and the small blueshift between the \oi{} 6300 and the  CaI 6439\,\AA{} line is just due to small differences in the dispersion solutions of the two echelle orders.\label{fig:oi_all}
}
\end{figure}

Optical spectra from T Tauri stars are known to have strong forbidden emission lines from [OI] at 6300, 6364, and 5577\,\AA{} \citep{hartigan95}. TW~Hya is no exception and these transitions are easily identified in all individual spectra. The strongest of all  is the [OI] line at 6300\,\AA{} with a median EW of 0.47\,\AA, which was found not to vary significantly
with time \citep{alencar02}. Following \citet{hartigan95} and assuming a mean R magnitude of 10.2, we convert this EW into a line
luminosity of 10$^{-5}$\,L$_\sun$. The optical spectrum of TW~Hya is known to vary, the most detailed V magnitude monitoring reports
night-to-night variations of $\Delta V \simeq 0.35$ (seasonal changes have smaller amplitudes, \citealt{rucinski08}). The same variations in R
convert into a [OI] luminosity range of 0.8-1.6$\times 10^{-5}$\,L$_\sun$. The 6364\,\AA{} [OI] line is a factor of $\sim$3 weaker than the [OI]
6300\,\AA{}, as expected from the ratio of the Einstein coefficients, while the 5577\,\AA{} [OI] line is a factor of $\sim$7 weaker on average (see Table~\ref{tab:optlines}). 
 These lines are also spectrally resolved with FEROS, the mean (and median) FWHM of the \oi{} 6300\,\AA{} is 11.5$\pm$0.4\,km/s
or an intrinsic line width of $\sim$10\,km/s after deconvolution with the 6\,km/s instrumental FWHM. The other weaker [OI] lines have similar
FWHMs. As already reported in \citet{alencar02} the \oi{} 6300\AA{} line presents only a narrow component, is centered at the stellar rest
frame, and there is no indication of a blueshifted high-velocity component from jets/outflows.  We use 40 spectra from \citet{alencar02} to
further investigate whether there is a blueshift of just a few km/s in the \oi{} peak emission\footnote{We excluded the 10 spectra obtained by \citet{alencar02} in December
1998 because the terrestrial night-sky emission line at [OI] 6300\AA{} falls at +12\,km/s in the stellocentric frame thus introducing a clear
bump on the red side of the \oi{} line from TW~Hya. For the other 40 spectra obtained in April and February the terrestrial \oi{} line is
blueshifted by 23\,km/s and is within 1\,km/s from the \oi{} line of TW~Hya respectively. Thus, the \oi{} line profiles from TW~Hya are not affected by
the night-sky emission in these 40 spectra}. To do that we rely not only on the absolute wavelength calibration performed previously but also
compare the location of the \oi{} 6300\,\AA{} line with respect to other known photospheric lines. Fig.~\ref{fig:oi_all} shows the distribution
of velocities for the \oi{} 6300\,\AA{} line in the stellocentric frame measured via the photospheric CaI 6439\,\AA{} line (black dashed line).
This histogram peaks at negative velocities but the median difference velocity is only -0.1\,km/s with a standard deviation as large as 0.15\,km/s. While this
tiny blueshift might indicate a flow, we show in the same figure that it is almost certainly related to small differences in the dispersion solution between
the two different echelle orders in which the \oi{} 6300\,\AA{} and the CaI 6363\,\AA{} fall. To demonstrate this we overplot the
difference between the  \oi{} 6300\,\AA{} and the \oi{} 6364\,\AA{} which falls in the same order as the CaI photospheric line and
should trace the same disk emission as the \oi{} 6300\,\AA{} line.  Fig.~\ref{fig:oi_all} clearly shows that the two velocity distributions are
identical, the median difference between the \oi{} 6300\,\AA{} and the \oi{} 6364\,\AA{} being -0.1\,km/s and standard deviation being
0.16\,km/s. These tests demonstrate that the [OI] optical lines are not blueshifted.



In addition to the oxygen lines, we also detect the [S~II] line at 4069\,\AA{} and the  Mg~I] line at 4571\,\AA{} with luminosities that are about $\sim$10 times lower than 
the [OI] 6300\,\AA{} luminosity (Table~\ref{tab:optlines}). The [S~II] transition at 4076\,\AA{} comes from very similar upper
energies ($\sim$3\,eV) as the 4069\,\AA{} transition but the Einstein coefficient is 2.5 times smaller, consistent with our non-detection and
3$\sigma$ upper limit.  The  [S~II] at 4069\,\AA{} and the Mg~I] line at 4571\,\AA{} peak at the stellar radial velocity as the [O~I] lines. 
Their luminosities are similar to those predicted by the X-EUV primordial disk model with Log(Lx)=29.3 (Ercolano \& Owen 2010). However, we note that this Xray luminosity is low for TW~Hya, about an order of magnitude lower than what has been estimated by e.g. Kastner et al. (1999). In the framework of photoevaporative disk models, the most puzzling result is the non-detection of the other two [S~II] transitions at
6716 and 6731\,\AA{} with stringent upper limits pointing to fluxes $\sim$50 times lower than the [O~I] 6300\,\AA{} line. As mentioned in Sect.~\ref{Sect:PhotModels}, both EUV and Xray irradiated disks should produce a strong [S~II] doublet for gas at low density and flowing, similar in strength or just a factor of a few weaker than the  [OI] 6300\,\AA{} line. While the [SII] 6716/6731 doublet should trace more extended low density gas than other forbidden lines due to its much lower critical density\footnote{according to EUV models the radial extension is a factor of 2 greater than the \neii{} emission (out to about $\pm$20\,AU from the star) and less concentrated to the center}, it is unlikely that the FEROS spectra lost most of the [SII] 6716/6731 flux since the fiber diameter projected on the sky is  2.7$''$ and covers radii out to $\sim$70\,AU at the distance of TW~Hya.  Thus, the non-detection of these transitions points to specific conditions in the disk emitting region, specifically to dense gas, as we shall discuss later.


Finally, the centrally peaked profiles and the FWHM of the optical lines can help constrain the extension of the emitting region. The [OI]
6300\,\AA{} line has an intrinsic width of 10\,km/s, the [SII] and MgI] lines are similarly narrow $\sim$10-12\,km/s, hence likely trace
similar disk radii. For a given ring of gas in Keplerian rotation extending out to a radius $r_{\rm o}$ from the star the absence of a
double peak implies: $2\,\sqrt{(\frac{GM_{\star}}{r{\rm o}})} sin(i) < \Delta v_{\rm FEROS}$ or $r_{\rm o} >$0.4-1\,AU for the mass and disk
inclinations reported for TW~Hya (Table~\ref{tab:twhyavalues}). Again for gas in Keplerian rotation, the inner ring radius $r_{\rm i}$
determines the maximum velocity in the line profile. Because the observed [OI] 6300\,\AA{} profiles extend out to $\pm$10\,km/s (see
Sect.~\ref{Sect:profiles}) the inner radius of the gas ring should be $<$0.2\,AU and can be as close to the star as 0.06\,AU for a 4\degr{} disk
inclination. This inner disk gas is likely not photoevaporating. However, if most of the emission is extended beyond about 1\,AU and gas is
photoevaporating, detailed disk models are necessary to compute line profiles and  estimate the extension of the emitting region (see
Sect.~\ref{sect:neutralwind}).

\subsection{Comparison of [Ne~II] and [O~I] line profiles}\label{Sect:profiles}

Our first high-resolution VISIR spectrum of TW~Hya showed a hint for an asymmetry in the \neii{} line
profile (see Fig.~4 from \citealt{pascucci09}),
with possibly more emission on the blue side than on the red side of the line. A flux enhancement on the blue side was also pointed out by
\citet{herczeg07} in their lower signal-to-noise spectrum. Such an enhancement is expected for face-on disks with an optically thick midplane
blocking the view of the redshifted outflowing gas  (\citealt{alexander08,eo10}). We use the 8 additional spectroscopic
exposures acquired in this observational campaign to further investigate the \neii{} 12.81\,\micron{} profile. To check for asymmetries we apply the
following method: We produce for each line a mirror spectrum (in velocity scale), then we apply small shifts in velocities to the
mirror spectrum such that the upper half of the lines matches best, finally we subtract the shifted-mirror and original spectra. This method is
especially sensitive to detect any blue or red excess emission in the wings of the line compared to its core emission and is more suited to lines that are poorly sampled in velocity than the line bisector method (e.g., \citealt{gray80}).

\begin{figure}
\includegraphics[scale=0.8]{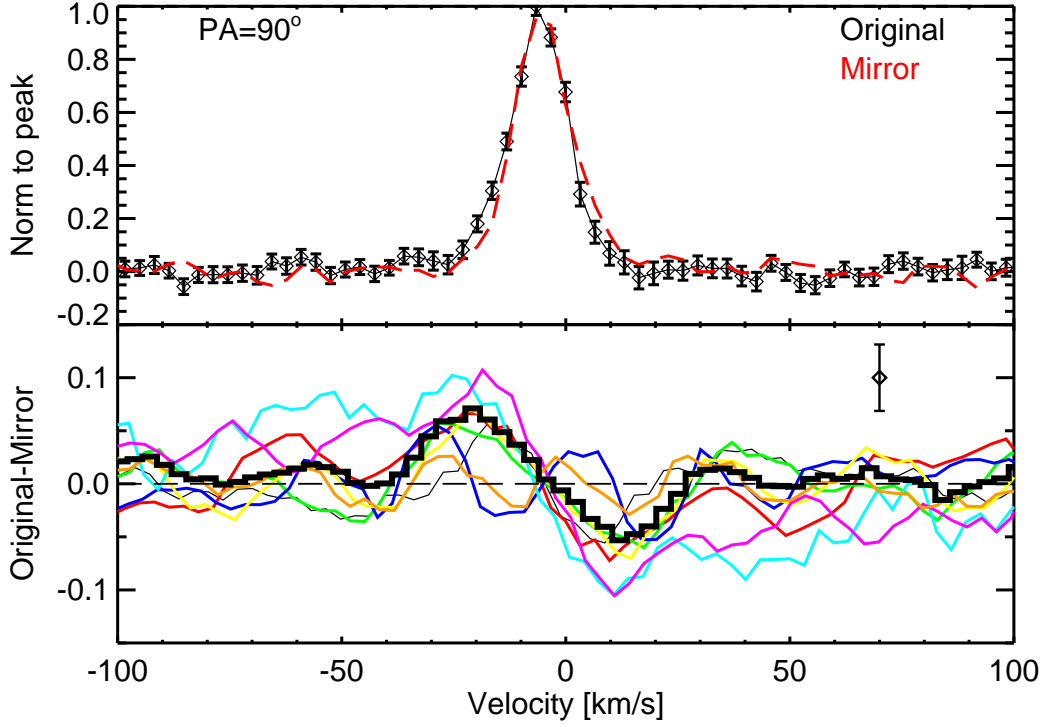}
\caption{\neii{} line profiles from TW~Hya. The upper panel gives the original (black) and
mirror-shifted (red) profile for PA=90\degr .  Note that the continuum is subtracted and the emission is normalized to the peak.
The lower panel shows the difference curves (original-mirror) for all eight PAs. The color coding is as follows: light black
for PA 180\degr , red for 90\degr , green for 0\degr , blue for 270\degr , yellow for 153\degr , cyan for 63\degr , violet
for 243\degr , and orange for 333\degr. The average difference curve is overplotted using a thick black line. Note that most
difference curves (as well as the average difference curve) have an excess on the blue side, peaking at about -20\,km/s. This
shows that the line is asymmetric with a blue wing due to photoevaporating gas accelerating toward the observer.\label{fig:wings_atmcorrect} }
\end{figure}

\begin{figure}
 \includegraphics[scale=0.8]{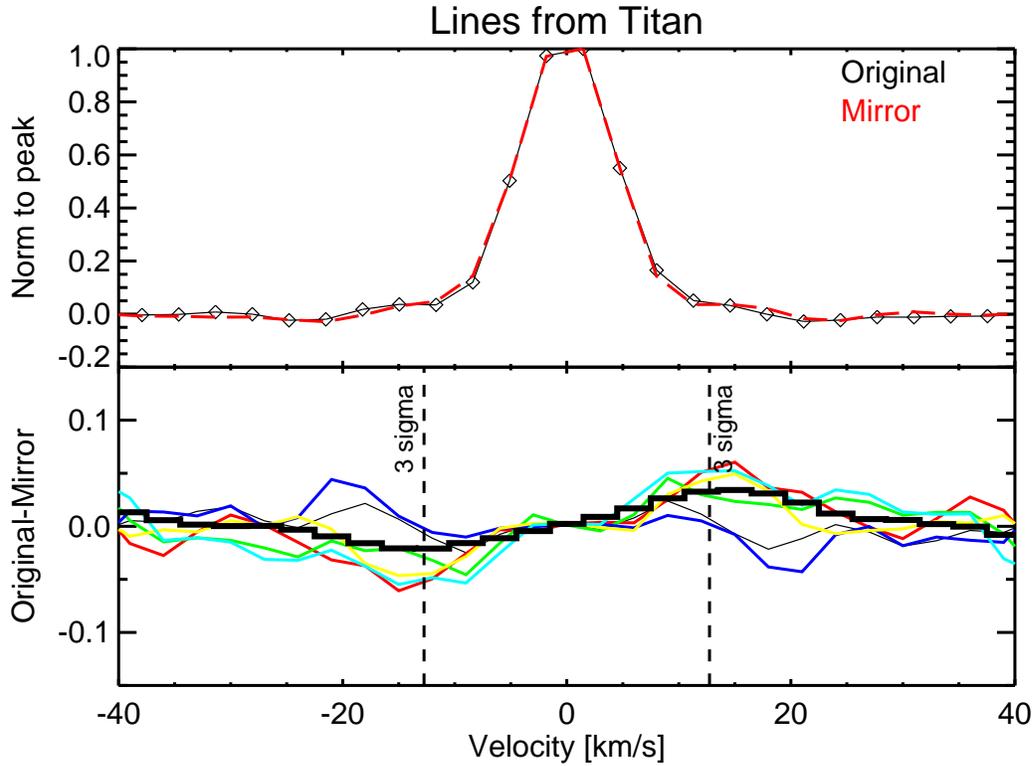}   
\caption{Line profiles from Titan  (PID 083.C-0883). The continuum near the line is removed and lines are normalized to the peak emission as for TW~Hya (also note the same y scale as in Fig.~\ref{fig:wings_atmcorrect}). Lower panel: difference  curves (original-mirror profiles) for 3 emission lines from two different spectra of Titan. This figure shows that the asymmetry seen in the \neii{} line profiles of TW~Hya is not an instrumental artifact.\label{fig:std_ettore}
}
\end{figure}

Fig.~\ref{fig:wings_atmcorrect} summarizes our findings. Individual line profiles as well as the median of line profiles (thick black line) show
an excess emission on the blue side of the \neii{} line peaking around -20\,km/s and a typical shift of the peak centroid of -5\,km/s\footnote{The two most different curves are for PA=270 and 333\degr. The line profile from the first PA shows decreased emission (one pixel) at $\sim$20\,km/s
and a positive (one pixel) emission at $\sim$-30\,km/s. The subtraction of the mirror-shifted profile (having reduced
emission at -30\,km/s and excess at 20\,km/s) generates the pronounced peak at -30\,km/s in the original-mirror profile. The
line for PA=333\degr{} is consistent with a symmetric profile.}. The blue excess is also visible as decreased emission on the red side of the line at $\sim$10\,km/s (see upper and lower panels of Fig.~\ref{fig:wings_atmcorrect}). We can confirm that this asymmetry is
intrinsic to the source emission and not an instrumental effect by applying the same method to spectrally unresolved
emission lines from Titan (see Fig.~\ref{fig:std_ettore}).

We then apply the same method of mirrored spectra to simulated \neii{} line
profiles from photoevaporative disk winds. As explained in Sect.~\ref{Sect:PhotModels}, such profiles
are available for the EUV-only \citep{alexander08} and the X-EUV \citep{eo10} models.
These models use input parameters such as stellar
mass, Xray and EUV luminosities that are similar to measured values for
the TW~Hya system but were not specifically tuened for TW~Hya (in the case of the X-EUV we use the models with Log(L$_{\rm X}$)=30.3 which are appropriate for TW~Hya).
In all cases the disk midplane is optically thick thus introducing an asymmetry in the
line profile which is, as expected, more pronounced the closest the disk
is to the face-on orientation (see Fig.~\ref{fig:models}). Applying the
KS test to the difference curves shows that the XEUV 10\degr{} disk
inclination model has just a 6\% probability that the observed and model
profiles are drawn from the same parent population hinting to a
difference. All other models (EUV with $i=$3, 10\degr{} and XEUV with
$i=$3\degr) have probabilities as high as 40\% meaning that they are all equally
good matches to the data. The important result from this analysis is that
the \neii{} emission must arise from a region in the disk that is
optically thick to the redshifted outflowing gas. We shall discuss in
Sect.~\ref{Sect:ionized} what this implies for the extension of the
ionized wind component of the photoveporative flow.

\begin{figure}
\includegraphics[scale=0.8]{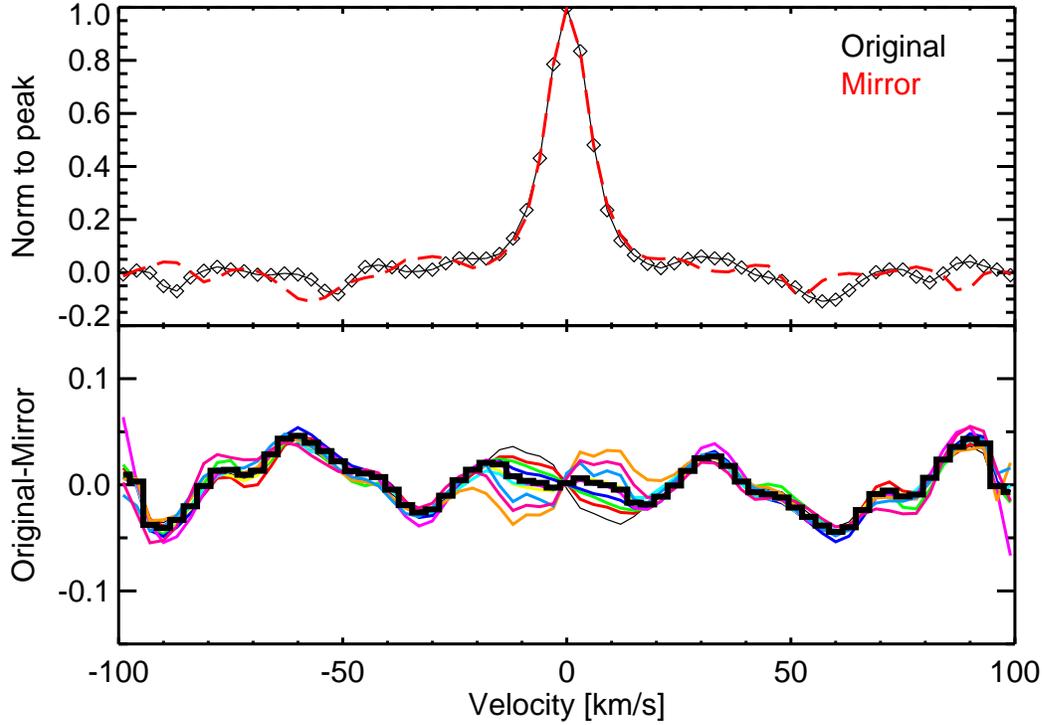}
\caption{Asymmetry analysis for the \oi{} 6300\AA{} emission lines of TW~Hya. The upper panel gives the original (black) and mirror-shifted (red) profile from one of the optical spectra obtained by \citet{alencar02}. The lower panel shows the difference curves (original-mirror) from the 10 spectra obtained in February 2000 \citep{alencar02}, different colors are for spectra obtained in different days. Note that unlike the \neii{} line at 12.81\micron{} the \oi{} line at 6300\AA{} is not blueshifted and has a symmetric profile. \label{fig:oi}
}
\end{figure}

\begin{figure}
\includegraphics[scale=0.8]{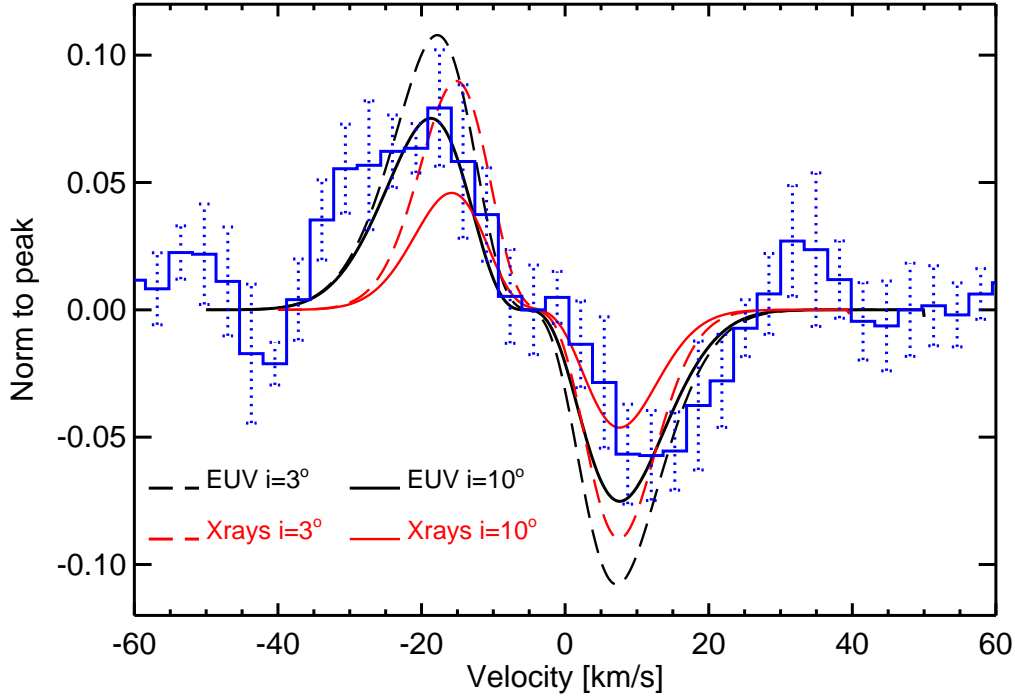}
\caption{Comparison of difference curves  (original-mirror profiles) for predicted and observed \neii{} line profiles. Black solid and dashed lines are the difference curves from the EUV photoevaporation model, while red solid and dashed lines are from the Xray photoevaporation model for disks inclined by 10 and 3\degr{} respectively. All models assume that the disk midplane  is optically thick. The blue line is the median of the difference curves from the observed \neii{} profiles of TW~Hya (see also Fig.~\ref{fig:wings_atmcorrect}) with the standard deviation of the curves overplotted.\label{fig:models}
}
\end{figure}

Finally, we also apply the method of mirrored spectra to the strongest of the optical lines, the [OI] 6300\,\AA{} line, to check for asymmetries in the
line profile. Fig.~\ref{fig:oi} shows that, unlike the \neii{} line, the [OI] profile is symmetric and the line is centered  at the stellar
velocity. As we will discuss in more detail in Sect.~\ref{Sect:ionized}, this profile could result from: i) pure Keplerian motion of gas bound to the disk (i.e. the \oi{} is not tracing the neutral component of the wind, it is not flowing); or ii)  a photoevaporative flow from a disk region that is optically thin to the redshifted outflowing gas. Examples of symmetric lines from photoevaporative flows are presented in  \citet{eo10} for disks with inner holes in their dust and gas distribution and emission confined within the hole. Because most of the opacity is expected to be from dust (see Sect.~\ref{Sect:ionized}) and dust extinction decreases with wavelength, our line profile analysis demonstrates that if scenario ii) applies the neutral \oi{} gas does not trace the same radii as the \neii{} emitting gas.

\subsection{Spectroastrometric signal of  a photoevaporative wind}\label{Sect:spectroastro}

We attempted for the first time spectroastrometry in the mid-infrared to  improve upon the spatial
resolution achievable with the VLT/VISIR and resolve the innermost region of the
photoevaporative wind. Spectroastrometry relies on the fact that the measurement of the centroid of a
flux distribution can be determined to much better than a pixel precision depending on the
signal-to-noise of the observations. In the case of VISIR one pixel corresponds to 127\,mas meaning
 that we could investigate the \neii{} emission at scales better than 6\,AU at the distance of TW~Hya.
Asymmetries in the emission at a specific wavelength appear as
shifts in the centroid of the flux distribution. Using different slit PAs enables to determine the
orientations at which asymmetries are stronger and thus help constraining the distribution of the source
emission.

\begin{figure}[h!]
\begin{center}
\mbox{ \subfigure{\includegraphics[scale=0.5]{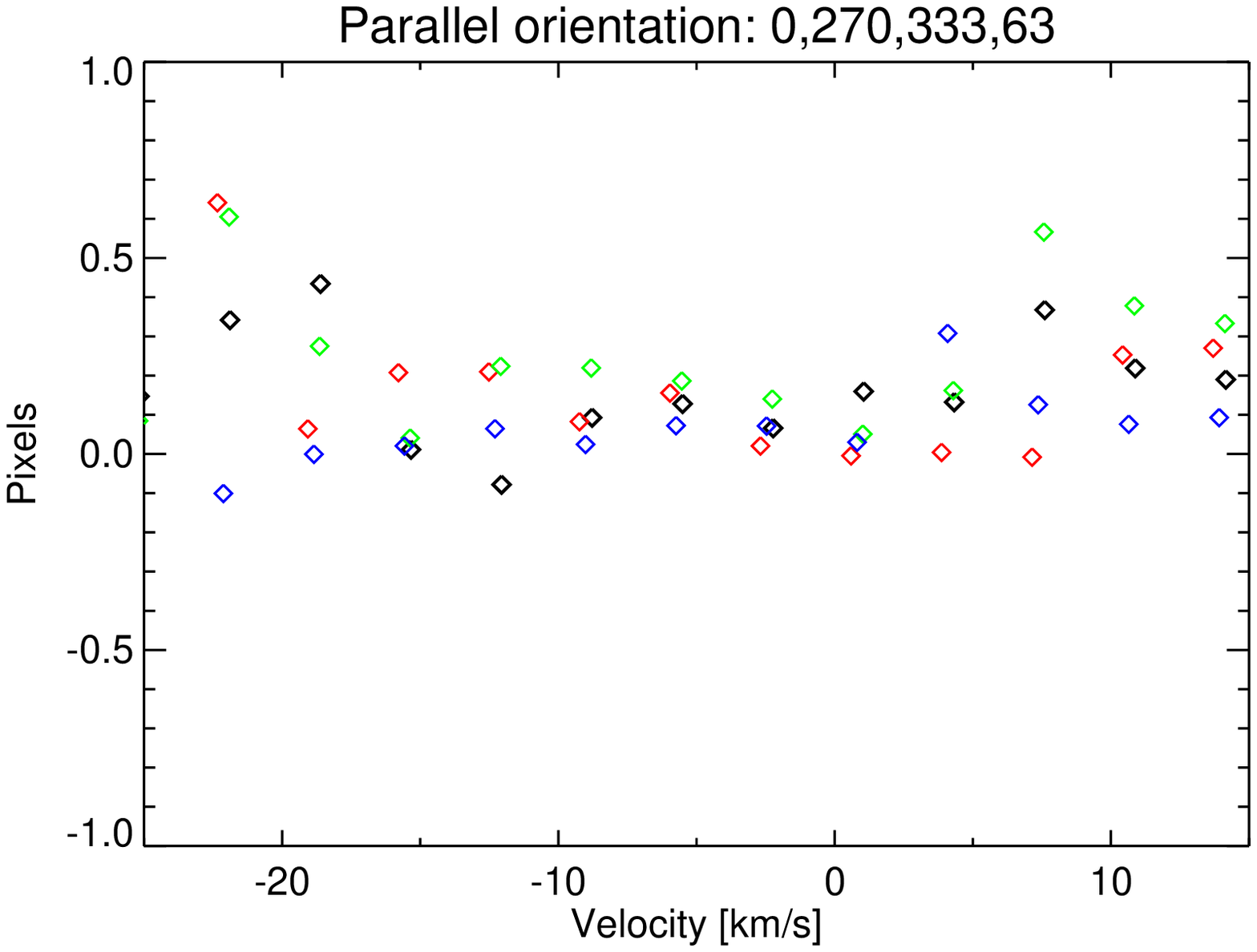}}\quad
\subfigure{\includegraphics[scale=0.5]{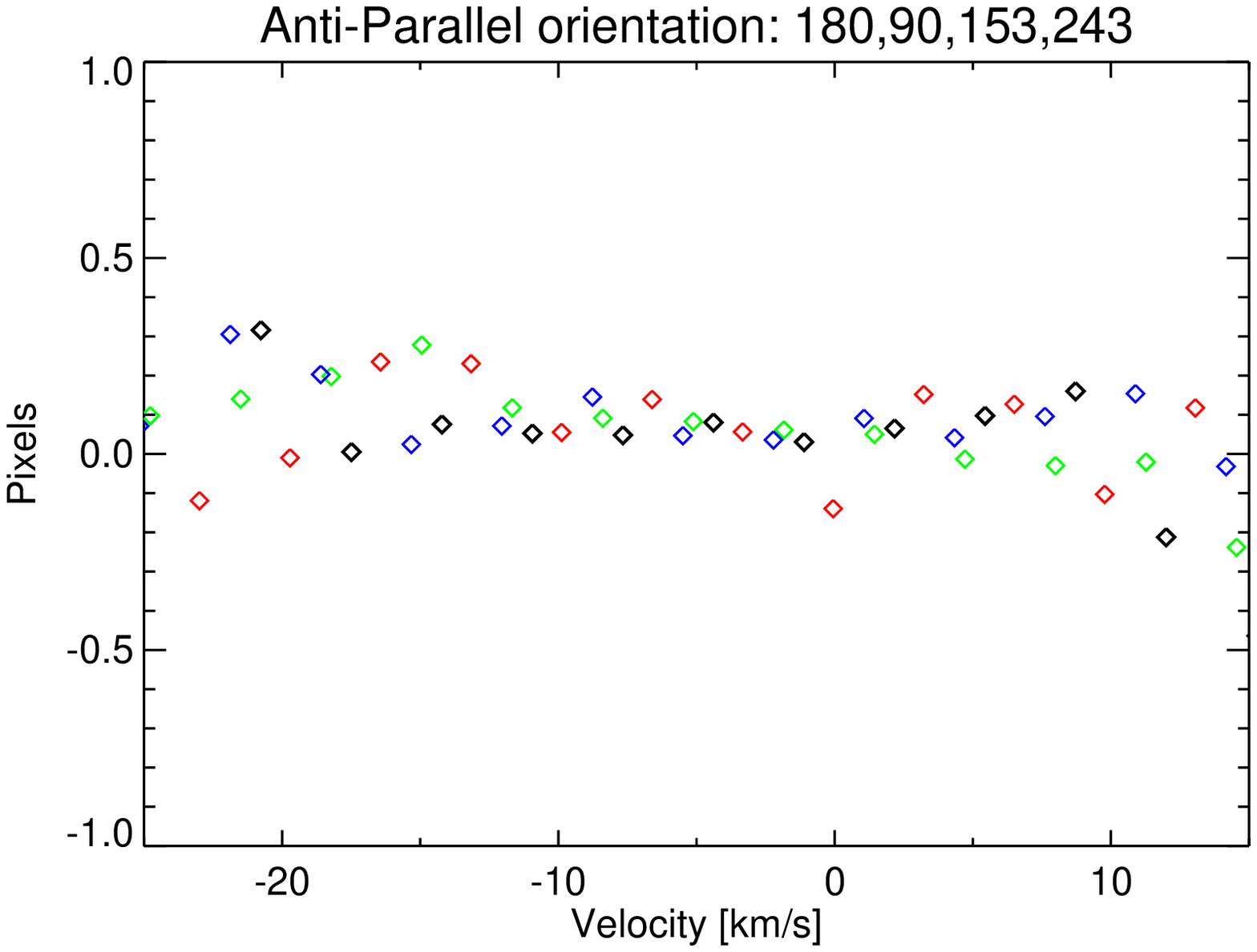}} }
\caption{Centroid in pixels versus velocities in the stellocentric frame for the parallel (left panel) and anti-parallel (right panel) slit orientations. The color coding is as follows: black for 0 and 180\degr , red for 270 and 90\degr , green for 333 and 153\degr , and blue for 63 and 243\degr .\label{fig:paantipa}}
\end{center}
\end{figure}

To retrieve the spectroastrometric signal from our VLT/VISIR data we extract spatial profiles in each
velocity channel and measure the spatial centroid of the \neii{} emission in cross-dispersion direction
with respect to the centroid of the emission in the continuum. The continuum is measured as the mean of
the continuum centroids on the red and on the blue side of the \neii{} line. Because of the low S/N on
the continuum we summed up the spatial profiles over more than 10 pixels in velocity, we have tested that
our results do not vary if we change the size of the velocity bins or the location where we measure the
continuum. To determine the centroid of the spatial profiles we have tested Gaussian and Lorentzian
profiles as well as the weighted mean over the FWHM  of VISIR in spatial direction as proposed by
\citet{pontoppidan08}. We find that fitting the emission with Lorentzian and Gaussian profiles provide
the same spectroastrometric signal and a more robust centroid than the weighted mean approach. As
uncertainty on the centroid we take the inverse of the S/N for each spatial profile where the noise is
measured as the standard deviation of the continuum outside the profile. This error reflects that
profiles extracted at velocities sampling the wings of the line have lower S/N, hence the centroid is
less well determined. The centroid is also diluted by the continuum flux at each velocity to take into
account the different continuum-to-line ratio moving from the core to the wing of the line (e.g., \citealt{wg08}). 
Fig.~\ref{fig:paantipa} shows the centroids in pixels versus velocities in the stellocentric frame for the parallel and anti-parallel slit orientations separately. 
This way we can better search for any possible instrumental effects on the centroids because such effects would appear as shifts of the same sign and intensity in the couple of parallel and anti-parallel slits \citep{bailey98}. The plot does not show any strong similar effect but small amplitude modulations across the \neii{} line and nearby continuum. As we will see next the centroids are measured to an accuracy of a tenth of a pixel close to the \neii{} peak emission and typically to half of a pixel in the wings where the S/N of the flux distribution in cross-dispersion direction is low\footnote{We note that the low accuracy in the centroid for PA=63\degr{} is due to the lower flux in the line and continuum likely caused by the fact that the source was not well centered in the slit for this PA, see Fig.~1}.

\subsubsection{Comparison between observations and EUV-model predictions}\label{Sect:modelwind}

The simple EUV-only disk model is computationally fast enough that we can create channel maps for the \neii{} emission and compute spectroastrometric curves for the slit orientations used in this observational campaign.  The model is essentially identical to the EUV-only model used by \citet{font04} and \citet{alexander08} with a radially continuous gas distribution.  The only differences between our hydrodynamic model and that of \citet{alexander08} are the resolution of the model grid, and the location of the lower boundary.  As we wish to generate spatially-resolved emission maps, it is necessary to use higher spatial resolution than that used by \citet{alexander08}.  In addition, we now assume that the underlying disk has a finite thickness, and place the base of the ionized wind at $z/r \simeq 0.15$ (e.g., \citealt{hollenbach94}). This has only a small effect on the predicted line profile, reducing the peak blue-shift by approximately 1\,km/s.  Our polar [$(r,\theta)$] computational grid covers the range $r = [0.03r_{\mathrm g},10.0r_{\mathrm g}]$, $\theta = [0,75^{\circ}]$, with grid cells that are logarithmically spaced in $r$ and linearly spaced in $\theta$.  We use $N_r = 1113$ cells in the radial direction and $N_{\theta} = 250$ cells in the polar direction. When the grid is expanded to three dimensions (by assuming reflective symmetry about the disk mid-plane and azimuthal symmetry around the polar axis), the resulting grid has $1113\times 600 \times 1200$ cells in the $r$, $\theta$ and $\phi$ directions respectively.

We compute the line emission from each grid cell in the same manner as \citet{alexander08}.  However, following the results of \citet{hollenbach09} we adopt an ionization fraction for Ne\,{\sc ii}  of 0.75 (rather than the somewhat arbitrary value of 0.33 used previously).  For comparison to the TW~Hya system we adopt a stellar mass of $M_* = 0.7$\,M$_\sun$ (thus $r_{\mathrm g} = 6.2\,$AU) and a disk inclination of $i=4^{\circ}$ (see Table~\ref{tab:twhyavalues} for references).  We set the ionizing flux $\Phi$ by matching the predicted line luminosity to the observed value of $3.1\times10^{-6}$L$_{\odot}$: we find that $\Phi = 7.5\times 10^{40}$ ionizing photons per second matches the observed line flux to $\sim 1$\% accuracy.  The integrated mass-loss rate from the disk is $\sim 1.5 \times 10^{-10}$\,M$_\sun$.  We create an integrated line profile (identical to that of \citealt{alexander08}) by summing the contributions from every cell on the grid, and also generate spatially resolved emission maps by de-projecting the three-dimensional grid on to the sky plane.  This is done at each (spectral) velocity to generate a three-dimensional data-cube.  The model data cube has 201x201 pixels which are 0.255\,AU on a side (equivalent to 5\,mas at the 51\,pc distance of TW Hya), and a velocity resolution of 0.1\,km/s.  We model the spectral resolution of VISIR by convolving the data cube (in the velocity direction) with a Gaussian profile with half-width $\sigma = 5$\,km/s, to give a simulated spectral resolving power of $R = 30,000$.  

\begin{figure}
\includegraphics[scale=0.8,angle=90]{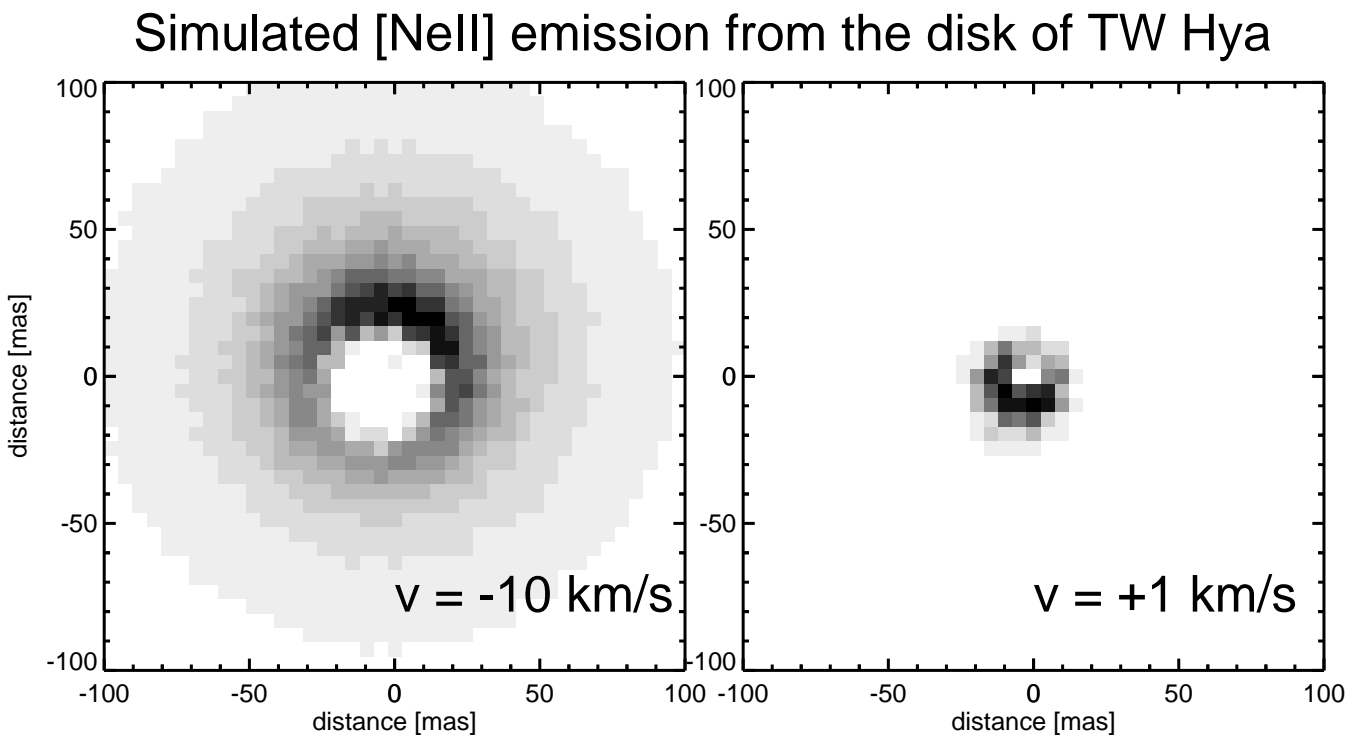}
\caption{Modelled \neii{} emission for two velocity channels, -10\,km/s and +1\,km/s. The asymmetry is mainly due to the combination of disk inclination (4\degr) and Keplerian
rotation. The wind is mainly symmetric but introduces deviations from the spectroastrometric Keplerian profiles, see text. The TW~Hya disk position angle is 333\degr. On the left panel the black rim N-W is a factor of $\sim$3 brighter than the grey emission S-E. On the right panel the contrast between the bright emission (now S-E) and the weaker emission (N-W) is about 5 but occurs at a much smaller spatial scale. These images are convolved with the VISIR PSF when simulating the spectro-astrometric signal.\label{fig:Hughes}} 
\end{figure}

\begin{figure}
\includegraphics[scale=0.8]{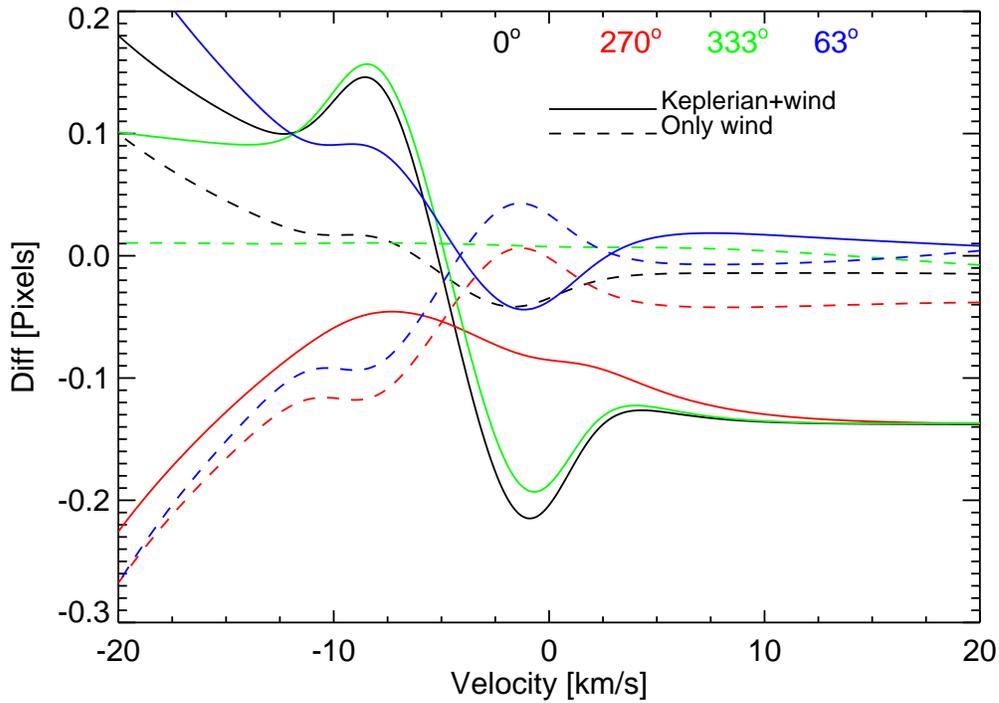}
\caption{Spectroastrometric signal (parallel minus antiparallel centroids) predicted by a disk photoevaporated by stellar EUV photons (solid lines). Dashed lines show only the wind component, i.e. when Keplerian rotation is artificially set to zero. The plot shows that the largest spectroastrometric signal from the disk wind is expected for PA=63\degr, the slit orientation perpendicular to the disk major axis. One pixel on VISIR is 127\,mas.\label{fig:spectroastro_model}
}
\end{figure}

Fig.~\ref{fig:Hughes} shows two simulated images of the \neii{} emission taken at a negative and a positive velocity, the image orientation is as in \citet{hughes10} when N is up and left is E in the figure.
To simulate the spectroastrometric signal we apply a slit in the x-axis equal to the VISIR slit width of 400\,mas, collapse the flux in the x-axis and evaluate
the centroid of the spatial profile in the y-axis after convolution with the VISIR PSF (FWHM of 300\,mas) for each velocity channel. To
simulate the rotation of the slit we rotate the simulated images by the corresponding angle and then apply the slit  as described
above. Fig.~\ref{fig:spectroastro_model} shows the expected spectroastrometric signal as parallel minus anti-parallel centroids in VISIR pixels
 for the photoevaporative disk model (solid lines) and for the wind only, i.e. when Keplerian rotation is set artificially to zero (dashed lines). One sees that for a slit orientation close to the disk major axis (PA=333\degr) the spectroastrometric signal is dominated by the Keplerian component of the gas, the asymmetry in the emission is small but present due to the small but non-zero disk inclination. The positive displacement at negative velocities can be understood by looking at Fig.~\ref{fig:Hughes}: the emission at negative velocities is stronger in the direction of PA=333\degr{} and slit orientations close to it. The wind component dominates the signal for the slit orientation perpendicular to the disk major axis (63\degr{} but it is prominent also for 270\degr) because at that orientation the Keplerian component is close to zero. The main results from the modeling are as follows: i) the expected shift in centroid is small, about 0.2 pixels, and ii) the shift is primarily set by projection effects (disk inclination) and the Keplerian component of the gas since the wind is more symmetrically distributed around the star.

\begin{figure}
\includegraphics[scale=0.8]{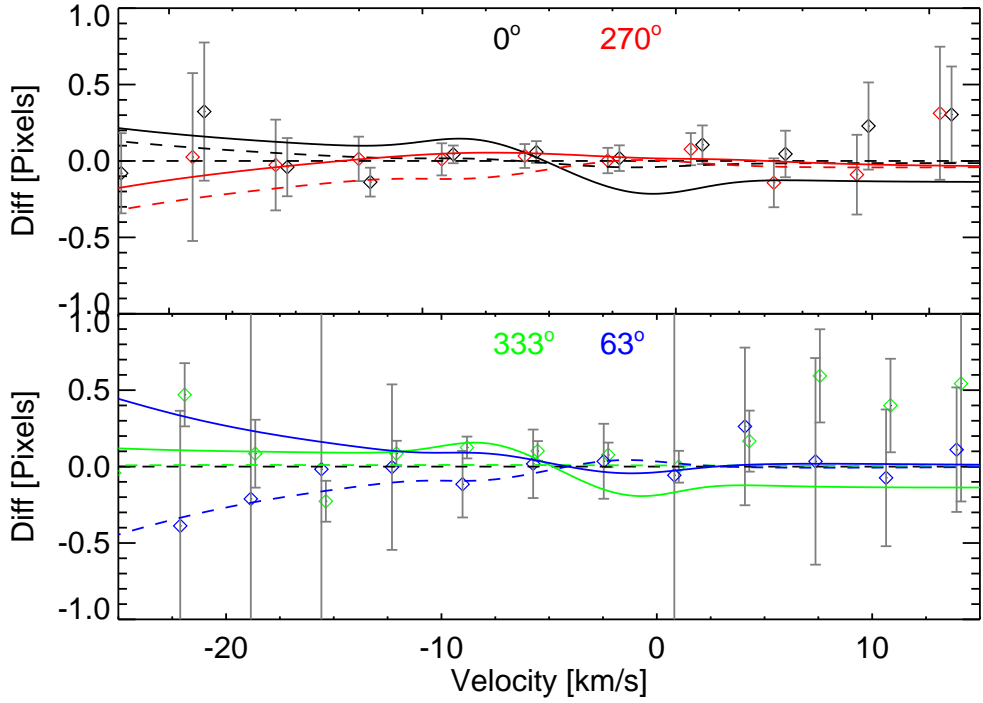}
\caption{Spectroastrometric signal observed (diamonds) and predicted (lines, see Fig.~\ref{fig:spectroastro_model} for color coding). The y-axis gives the difference in pixels between a position angle and its antiparallel position angle (e.g., 0\degr corresponds to 0\degr-180\degr). 
One pixel on VISIR is 127\,mas, therefore the level of asymmetry we detect has a radial extension of $\sim$1.3\,AU.\label{fig:spectroastro}
}
\end{figure}

To compare model predictions with observations, we subtract the measured anti-parallel centroid positions from the parallel positions (after interpolation over the same velocity grid), see Fig.~\ref{fig:spectroastro}. For PAs 270 and 63\degr{}, slits oriented close to the disk minor axis, the spectroastrometric shift is less than 0.1\,pixel as expected. For the other two slit orientations, sampling the disk major axis, we do not detect the S-shape profile predicted for gas in Keplerian rotation. We see a hint of such a profile only for PA=333\degr{} negative velocities which is however not present at the positive velocities. A smaller shift than predicted could result from a \neii{} emission that is more symmetric than predicted, from a smaller disk inclination, and/or from \neii{} emission more closely concentrated to the central star (in this last case, the convolution with the VISIR PSF would smear out the contrast between the bright and faint rims of the \neii{} emission). Further observations at even higher spectral resolution and sensitivity, especially in the wings of the line, are necessary to confirm these first results and to possibly identify the small spectroastrometric shift from photoevaporating gas that is expected for slit orientations along the minor axis.

\section{Physical implications on the disk wind}\label{Sect:ionized}

Before discussing the possible wind structure from the disk of TW~Hya we wish to briefly summarize the main observational results presented in the previous Sections. 
\begin{itemize}
\item We confirm that the \neii{} emission line at 12.81\,\micron{} is blueshifted by $\sim$5\,km/s with respect to the stellar velocity. The line is relatively narrow (FWHM$\sim$13\,km/s) and the profile is asymmetric, with more emission on the blue than on the red side.
\item None of the forbidden and semi-permitted optical lines investigated here appear blueshifted. On the contrary, these lines are centered in the stellocentric frame to within about 0.15\,km/s. The line profiles are symmetric and the similar FWHMs ($\sim$10\,km/s) from different transitions point to optical lines tracing gas at similar radii.
\item The very small or zero spectroastrometric shifts for the observed PAs suggest that the \neii{} emission from the photoevaporative wind is predominantly spatially symmetric.  
\end{itemize}

What are these results telling us about the structure of the photoevaporative wind and its extension?

\subsection{The unbound disk component}\label{Sect:unbound}

The small blueshift in the \neii{} line is a clear tell-tale sign of photoevaporating gas for a face-on configuration like that of TW~Hya. Hence, we are sure that this line is probing the ionized component of the disk wind. This component could be the dominant wind component in the EUV case or a trace component of the mostly neutral wind in the Xray case.
Our observations constrain the radial extension of the ionized wind component via the \neii{} line. By recovering all flux measured with {\it Spitzer}/IRS, we demonstrate that the \neii{} emission arises within 10\,AU from the star. We note that although the radial extension of the mass loss is very different for pure EUV and X-EUV irradiation, with Xrays producing 50\% of the mass loss at tens of AU \citep{owen10}, the \neii{} luminosity decreases steeply with radial distance  from the star (as $\sim 1/r$) hence the emission is dominated by the flux from small radii. For the specific EUV and X-EUV models discussed here 90\% of the \neii{} emission arises within 12\,AU and 15\,AU respectively, which is consistent with the 10\,AU radial extension inferred observationally. 

The blue excess in the wing of the \neii{} line  adds a further constraint on the radii traced by this wind component. We have shown in Sect.~\ref{Sect:profiles} that primordial disks (no gaps) irradiated by EUV and Xrays can reproduce the \neii{}  line profile and peak emission.  Here we ask the question: how much \neii{} emission could originate in an optically thin region and still be consistent with the observed \neii{} profile? To simulate the \neii{} profile  from optically thin disk radii  we take a gaussian centered at zero velocity and FWHM set by thermal broadening for gas at $\sim$5,000\,K (Xray heating) or $\sim$10,000\,K (EUV heating) and convolved with the FWHM of the VISIR spectrograph. We add as much as 10, 20, and 30\% of the total emission to the \neii{} profile from a fully thick disk and compare the resulting profile with the observed one. We find that already a 10\% contribution from radii optically thin to the \neii{} emission shifts the line peak emission redward to $\sim$-4\,km/s, marginally consistent with our measurements, and increases the symmetry of the line. This simple approach shows that more than 80\% of the \neii{} emission should come from a disk region that is fully optically thick to the \neii{} line. What is then the source of opacity within 10\,AU from the central star? This is a relevant question for transition-type sources like TW~Hya whose SEDs clearly point to opacity gaps in their disks.

First we look at the gas component and show that the \neii{} line is optically thin and gas alone cannot be responsible for the blueshift and line asymmetry. \citet{hm89} calculated that a column of Ne$^+$ of a few times $10^{18}$\,cm$^{-2}$ (or a few $10^{22}$ H atoms for a solar Ne/H abundance ratio) is necessary to produce an optical depth of 1 at the center of the \neii{} line. This density has to be compared with  $<10^{16}$\,cm$^{-2}$ Ne$^+$ in the EUV layer and $<10^{17}$\,cm$^{-2}$ in the Xray layer \citep{hollenbach09,glassgold07}, clearly not sufficient to make the line optically thick. That the line is optically thin is also clear from the observed line profile which does not show any absorption at the line center, where the optical depth is the highest.

Sub-micron dust grains are the next major source of opacity. According to the disk model of \citet{calvet02} the inner disk of TW~Hya could have only a tiny amount (0.5 lunar masses) of micron-sized grains out to $\sim$4\,AU and thus would be optically thin at visible and even more at infrared wavelengths out to that radial distance. However, \citet{ratzka07} show that this model gives a factor of 2 too high mid-infrared visibilities as well as the wrong wavelength dependence of the visibility. They argue that the dust inner rim, hence the transition from thin to thick disk, must happen much closer in at about 0.5-0.8\,AU. More recently, \citet{akeson} also argue for the presence of optically thick dust within 4\,AU. They propose a more complex structure with an optically thick dust ring centered at 0.5\,AU, a dust gap, and the full optically thick dust disk starting at $\sim$4\,AU. Our simple approach of combining \neii{} profiles from optically thin and thick disk radii suggests that more than 80\% of the \neii{} emission should arise from disk radii where the midplane is optically thick to the \neii{} emission. If the optically thin disk extends out to 4\,AU as in the model of \citet{calvet02}, then the observed \neii{} profile  cannot be reproduced by the EUV-only and X-EUV models from \citet{alexander08} and \citet{eo10} because in these models $\sim$50\% of the \neii{} emission arises within 4\,AU. If the dust inner rim is at $\sim$1\,AU, as in the \citet{ratzka07} model, then these models can explain the blueshift and asymmetry in the line profile since they both predict that $\sim$90\% of the \neii{} emission comes from outside 1\,AU, i.e. from disk radii where the midplane would be optically thick. A different model, also consistent with our observations, is that from  Gorti et al. (2011). From a detailed analysis of many gas emission lines from TW~Hya and using the Calvet et al. dust distribution with a {\it static} disk model, they find that 75\% of the \neii{} emission line comes from the surface of the optically thick dust disk beyond the 4\,AU gap.

\subsection{The disk component traced by optical lines}\label{sect:neutralwind}
Unlike in the case of the \neii{} line, the forbidden and semi-permitted optical lines investigated here are not blueshifted with respect to the stellar radial velocity. The lack of blueshift could have two origins: i) the optical lines trace a wind component at disk radii where the midplane is optically thin to the redshifted outflowing gas; OR ii) the optical lines do not trace the photoevaporative flow but rather  gas that is bound to the star/disk system. 
In reference to the [OI] optical lines, this means that we do not have an unambiguous diagnostic for the neutral wind component of the disk as we have for the ionized component. In the following, we discuss what other constraints we can place on the disk region traced by the optical lines and what observations could discriminate between the two scenarios presented above.

First of all, the similarity in the FWHM of the [OI] 5577, 6300, 6364\,\AA{}, the [SII] at 4068\,\AA{}, and the  MgI] at 4571\,\AA{} suggests that these lines trace similar disk radii. Based on the centrally peaked profiles and maximum observed velocities, we argue in Sect.~\ref{Sect:otherlines} that the emission could extend as close as 0.06\,AU to the star, where the disk may be magnetospherically truncated, to and possibly beyond 1\,AU. The observed [OI] 6300/5577\,\AA{} line ratio of $\sim$7 as well as the non-detection of the  [SII] doublet 6716/6730\,\AA{} point to a region of high temperature and electron density. For  gas in local thermal equilibrium (LTE) and optically thin emission we can write the [SII] 6716 and  4069\,\AA{} ratio as:
\begin{equation}
\frac{L_{4069}}{L_{6716}} = \frac{n(^2P)  A_{4069}  \Delta E_{4069}}{n(^2D)  A_{6716}  \Delta E_{6716}}
\end{equation}
$n(^2P)$ and $n(^2D)$ are the upper level populations which in LTE are linked as $n(^2P) / n(^2D) \sim 4/6 \times e^{-14,000/T}$. To explain a line flux ratio higher than 6 we thus require gas hotter than $\sim$3,000\,K. A similar argument applied to the [OI] 6300 and 5577\,\AA{} lines yields  a similarly high temperature (Gorti et al. 2011). Such a high temperature can be reached within a few AU from the star in a disk atmosphere irradiated by stellar Xray but LTE requires very high electron densities ($n_e\gtrsim10^8$cm$^{-3}$) which may not be present in the photoevaporative wind (Gorti et al. 2011). We turn now to the strong [OI] lines and on the neutral component of the disk wind to investigate the two scenarios proposed at the beginning of this section.

In the first scenario, the [O~I] emission is thermal and arises from the hot ($>$5,000\,K) disk surface heated by stellar Xrays.  \citet{eo10} have shown that 
line profiles can be symmetric even if the gas is flowing and any blueshift in the peak emission can be small,  a few tenth of km/s, if most of the emission arises within the dust depleted inner disk\footnote{As in the case of the \neii{} line sub-micron dust grains are the main source of opacity. For the [OI] 6300\,\AA{} line the H column density to reach optical depth of one  is as high as 2$\times10^{25}$ cm$^{-2}$.}.  For the specific case of TW~Hya an inner dust hole of 2\,AU would produce a blueshift in the [OI] line greater than 1\,km/s and slightly asymmetric profiles while a 4\,AU hole would give just a shift of 0.4\,km/s and a symmetric profile pointing to the flow traced by the [OI] emission extending well beyond 2\,AU. The FWHM of the line is in both models relatively large $\sim$20\,km/s, a factor of 2 larger than measured. The line broadening can be reduced if the emission in the flow is suppressed and the bound \oi{} thermal component arising from within a few AU  dominates the emission. This strongly points to the second scenario in which the [OI] lines are not tracing the neutral component of the photoevaporative wind in the disk of TW~Hya but rather the bound disk component well inside the gravitational radius.

A variant to the second scenario is that proposed by Gorti et al. (2011). In this model, which is specifically for TW~Hya, most of the [O~I] emission is not thermal but produced from dissociation of OH molecules by stellar FUV photons leading to O atoms in electronically excited states and their subsequent decay to the ground state. Gorti et al. (2011) find that this mechanism gives a 6300/5577\,\AA{} line ratio of $\sim$7, very similar to the observed one. Because the \oi{} emission comes from cooler gas closer to the disk midplane the emitting gas would not participate to the photoevaporative flow and hence [OI] lines would be narrower and symmetric around the stellar velocity. The [SII] line ratios are also reproduced in this model since most of the emission arises in the hot and dense gas close to the star within the dust depleted inner disk. 

A way to directly discriminate between bound and unbound oxygen gas is through spectroastrometry in the [OI] 6300\,\AA{} line. For gas in Keplerian rotation the spectroastrometric signal will be symmetric in velocity and the strongest for a slit oriented along the disk major axis, the signal will be zero for a slit perpendicular to the major axis (see, e.g. Fig.~\ref{fig:spectroastro_model}). The presence of a photoevaporative wind alters the symmetric spectroastrometric signal and produces a shift in the flux distribution versus wavelength even for a slit perpendicular to the disk major axis. 
With a combination of good seeing ($\sim$1\arcsec) and high signal to-noise a few milli-arcsecond accuracy can be attained via spectroastrometry at visible wavelengths (see e.g. Takami et al. 2003). Hence, detecting asymmetries around 1\,AU, which would produce shifts of $\sim$20\,mas, is well within the reach of this technique.


\subsection{On determining photoevaporation rates}\label{sect:testmodels}

Our study of TW~Hya represents the first attempt to identify and combine information from multiple gas lines at infrared and optical wavelengths that could trace the photoevaporative disk wind.
Here, we summarize the observables reproduced by different disk models as applied to TW~Hya and discuss how to nail down photoevaporation rates from protoplanetary disks.

The intensity of the \neii{} line at 12.81\,\micron{} can be reproduced by all models: in the EUV-only model ionized neon atoms trace the skin of the disk  while in the  X-EUV and XE-FUV models they trace a deeper, mostly neutral layer that participates to the flow.
Any asymmetry in the line profile tells us where most of the emission arises with respect to the optically thick dust disk. A better determination of the transition from thin to thick disk could potentially validate or invalidate the EUV/Xray photoevaporation models (see Sect.~\ref{Sect:unbound}). 
The strong  \oi{} line at 6300\,\AA{}  from TW~Hya cannot arise from the EUV layer because this layer is mostly ionized and produces only weak [O~I] emission. 
While the [O~I] peak emission from several T Tauri stars in the Hartigan et al.~(1995) sample presents a small blueshift pointing to a possible origin in the Xray wind, this is not the case for TW~Hya. In addition, the small FWHM of the [O~I] line also argues against most of the emission tracing the wind and rather points to bound gas close to the star. 
The luminosities of the [S~II] at 4068 and the Mg~I]  at 4571\,\AA{} can be reproduced by the X-EUV primordial disk model but with a stellar Xray luminosity that is 10 times weaker than that of TW~Hya (Table~1 from Ercolano \& Owen 2010).
The non-detection of the [S~II] doublet at 6717 and 6731 \AA{} is also puzzling in the frame of both the EUV and the X-EUV model. For our results to be consistent with these models,  sulfur should be depleted in the disk wind.  Depletion of more refractory elements such as silicon and aluminum has been previously proposed for the gas surrounding TW~Hya (e.g., Lamzin et al. 2004). 
Elemental abundances therefore add an additional uncertainty when comparing observed  line fluxes to model calculations.
Alternatively the XE-FUV model developed for TW~Hya explains the non-detections of the [S~II] doublet and proposes that the observed [S~II] line at 4069\,\AA{} arises from dense, bound gas in the inner disk (Gorti et al. 2011). Similarly to the forbidden optical lines, the CO line profiles and their spectroastrometric signal are consistent with bound gas in Keplerian rotation around TW~Hya (Pontoppidan et al. 2008).

More effort should be devoted in identifying gas lines that trace the predominantly neutral winds of XE-FUV and X-EUV irradiated disks both around TW~Hya as well as around other young stars. Establishing the presence of this component would point to flow rates 1-2 orders of magnitude  larger than the 10$^{-10}$\,M$_\sun$/yr predicted by EUV-only photoevaporation models. In addition, attempts should be made to empirically measure the ionizing flux impinging on the surface of protoplanetary disks via for instance line flux ratios from two successive stages of ionization of a given element or from volatile elements tracing the flow.

\section{Concluding remarks}
In the present work we have provided observational constraints to the photoevaporative wind from the disk of TW~Hya.  
We confirm our previous finding that the \neii{} 12.81\,\micron{} line traces the ionized component of the
wind from the disk of TW~Hya  \citep{pascucci09}. This would be the dominant wind component in the EUV case and a trace component in the predominantly neutral Xray wind.
From the blueshift and asymmetry in the line profile, we estimate that most ($>$80\%) of the \neii{} flux arises from disk radii where the midplane is optically thick to the emission, meaning beyond the 1 or 4\,AU dust gap inferred from other observations. We note that depending on the observations/disk modeling this gap may be not completely evacuated of dust \citep{calvet02,ratzka07,akeson}. We also show that the \neii{} emission is confined within 10\,AU from the star. Our spectroastrometry  in the \neii{} line suggests that the wind flux is rather spatially symmetric around the star but
the current dataset is not of quality high enough to provide additional constraints to the structure of the photoevaporative flow. 
We find that optical forbidden lines such as the strong \oi{} emission line at 6300\,\AA{} have remarkably different profiles from the \neii{} 12.81\,\micron{} line, in that they are centered at the stellar velocity and are symmetric. The only way these lines could trace the photoevaporative flow is if they arise from a disk region physically different from that traced by the \neii{} line, specifically from within the optically thin dust gap.  However, the small FWHM of the optical lines indicates that most of the gas is likely bound, which would imply gas very close to the star (inside the $\sim$1\,AU critical radius for photoevaporation) if the Keplerian velocity contributes most to the line broadening.

The above characteristics of the gaseous disk of TW~Hya can be used to speculate on the origin of its transition-like SED. One possibility for creating dust gaps in circumstellar disks and still account for accretion onto the star is via dynamical clearing by forming giant planets \citep{la06}. For the specific case of TW~Hya, partial dust clearing is required at most out to $\sim$4\,AU from the star \citep{calvet02}. 
Recent hydrodynamic simulations by \citet{zhu11} suggest that one giant planet might account for this moderate-size gap (see also Gorti et al. 2011) and still result in the measured accretion rate of $10^{-9}$\,M$_\sun$/yr \citep{muzerolle00,alencar02}.
Another possibility is that the gap is not opened by a planet but rather by photoevaporation and we are catching the star in the process of dispersing its disk. Recent XEUV-driven photoevaporation models predict that there is a non-negligible amount of time (2-3$\times 10^{5}$\,yr) during which photoevaporation can carve a  moderate-size gap, accretion onto the star is still detectable, and dust grains are cleared out under the action of dust drag \citep{owen11}. 
We would like to stress that the on-going photoevaporation from the disk of TW~Hya reported here does not exclude the existence of a giant planet and does not preclude that a forming giant planet has dynamically cleared the inner disk. Infact, a giant planet might expedite the onset of substantial (and detectable) photoevaporation by depleting the inner disk surface density which reduces the stellar accretion rate and exposes the region out of the rim with direct unattenuated flux from the star.

Besides detecting a planet at a few AU from TW~Hya, there are other observables that could help discriminating between a planet-induced and a photoevaporation-induced disk gap. If the putative giant planet is on an eccentric orbit, one might see variations in the dust disk structure and related dust emission on the Keplerian timescale of the planet, a few years. Regardless of the planet's eccentricity, dynamical clearing would result in a sharp transition in surface density at the outer gap edge which could be imaged with the  Atacama Large Millimeter Array. 





\acknowledgments
IP is pleased to acknowledge support from the National Science Foundation (NSF) through an
Astronomy \& Astrophysics research grant (AST0908479). IP thanks E. Flaccomio for making available
the VISIR spectra of Titan and K. Pontoppidan for valuable discussions. RDA acknowledges support from the Science \& Technology Facilities
Council (STFC) through an Advanced Fellowship (ST/G00711X/1). SHPA acknowledges support from
Fapemig  and CAPES, brazilian research agencies. This research used the ALICE High Performance
Computing Facility at the University of Leicester.  Some resources on ALICE form part of the DiRAC
Facility jointly funded by STFC and the Large Facilities Capital Fund of BIS.



{\it Facilities:} \facility{VLT (VISIR)}, \facility{1.5m~ESO (FEROS)}.

\clearpage




\clearpage




\clearpage

\begin{deluxetable}{lccccclcc}
\tabletypesize{\scriptsize}
\tablecaption{Summary of the TW~Hya observations. The observing time (in U.T.), the airmass, 
and the heliocentric radial velocity corrections (v$_{\rm helio}$) are given at the beginning and at the
end of each spectroscopic exposure. On-source exposure times for TW~Hya were set to 1\,h, for standard stars exposures varied between 240 and 360\,s. In the second night, we observed only one standard star for PA=333\degr .\label{table:log}}
\tablewidth{0pt}
\tablehead{
\colhead{PA\tablenotemark{a}} & \colhead{Date} & \colhead{U.T.}&  \colhead{Airmass} &\colhead{v$_{\rm helio}$}& \colhead{Calibrator} & \colhead{Airmass} \\
\colhead{(\degr)} & \colhead{yyyy-mm-dd} &\colhead{hh:mm}& \colhead{} &\colhead{(km/s)} &\colhead{} & \colhead{}
}
\startdata

0         &2010-02-23 &02:12/03:41& 1.3/1.1& 11.53/11.40 & HD~90957 &  1.3 \\
180   &2010-02-23 &03:54/05:23& 1.1/1.0& 11.38/11.23& HD~90957 &  1.0 \\ 
270      &2010-02-23 &06:27/07:54&1.0/1.2 &11.11/10.97 & HD~101666 & 1.0   \\ 
90    &2010-02-23 &08:25/09:42&1.3/1.5&10.93/10.84 &HD~101666  &   1.5 \\   
\hline
333    &2010-02-24 &02:45/04:15&1.2/1.1&11.11/10.97&HD~90957&1.2 \\
153  &2010-02-24 &04:28/05:58&1.0/1.0&10.95/10.79& -- & --\\
63    &2010-02-24 &06:24/07:51&1.0/1.2&10.74/10.70& -- & -- \\
243   &2010-02-24 &08:07/09:38&1.2/1.7&10.58/10.47& -- & -- \\
\enddata
\tablenotetext{a}{The position angle (PA) is measured from N toward E.}
\end{deluxetable}

\begin{deluxetable}{lccccc}
\tabletypesize{\scriptsize}
\tablecaption{Main properties of the TW~Hya star/disk system inferred from other studies and utilized in this paper.\label{tab:twhyavalues}}
\tablewidth{0pt}
\tablehead{
\colhead{Property} & \colhead{Value} & \colhead{Reference}\\
}
\startdata

Stellar mass         & 0.7\,M$_\sun$   & 1\\
Distance  & 51$\pm$4\,pc & 2 \\ 
Disk inclination      &  4.3$\pm$1\degr,7$\pm$1\degr,5-6\degr     & 3,4,5 \\ 
Disk PA    & 332$\pm$10\degr, 335$\pm$2\degr       & 3,5 \\ 
Stellar ionizing flux & 10$^{41}$\,phot/s & 6\\ 
\enddata
\tablerefs{
(1)  \citealt{muzerolle00};
(2)  \citealt{mamajek05};
(3) \citealt{pontoppidan08};
(4) \citealt{qi04};
(5) \citealt{hughes10};
(6) \citealt{pascucci09}
}
\end{deluxetable}

\begin{deluxetable}{lccccc}
\tabletypesize{\scriptsize}
\tablecaption{Main parameters of the \neii{} line and adjacent continuum for different position angles (PA).\label{tab:var}}
\tablewidth{0pt}
\tablehead{
\colhead{PA} & \colhead{EW} & \colhead{FWHM}&  \colhead{v$_{\rm peak}$} &\colhead{Line Flux} &\colhead{Continuum} \\
\colhead{(\degr)} & \colhead{(\AA)} &\colhead{(km/s)}& \colhead{(km/s)} &\colhead{($\times 10^{-14}$\uflux)}&\colhead{(Jy)} 
}
\startdata

0         &43$\pm$1        & 15.2$\pm$0.4 &-5.0$\pm$0.2 &2.6$\pm$0.1&0.3\\
180   &47.1$\pm$0.8 & 16.9$\pm$0.4 &-4.9$\pm$0.1 &6.5$\pm$0.2&0.8\\ 
270      &  44$\pm$1      & 15.0$\pm$0.4 &-4.8$\pm$0.2&4.4$\pm$0.1&0.6\\ 
90    & 36$\pm$1       & 15.8$\pm$0.5 &-5.9$\pm$0.2&3.0$\pm$0.1&0.5\\   
\hline
333     & 47$\pm$1        & 15.9$\pm$0.4  &-6.2$\pm$0.2&3.2$\pm$0.1&0.4\\
153  &47.7$\pm$0.8 &  16.6$\pm$0.3  &-3.9$\pm$0.3&--&--\\
63    & 30$\pm$1      &  17.5$\pm$0.7   &-7.5$\pm$0.3&--&-- \\
243   & 35$\pm$1     &   17.2$\pm$0.6  &-3.9$\pm$0.3&--&-- \\
\enddata
\tablecomments{Errors on the EWs are computed using a Monte Carlo approach: we added a normally distributed noise to the spectrum (following
the error at each wavelength) and give as error the standard deviation of the EW distribution of 1000 such spectra. The errors on the FWHM,
peak emission, and line flux are computed from the gaussian fit to the \neii{} line taking into account the uncertainty at each wavelength.
Note that the absolute radial velocity accuracy reachable with VISIR is $\sim$2\,km/s (Sect.~\ref{Sect:waveaccuracy}), larger than the uncertainties on
the \neii{} peak velocities derived from the gaussian fit.}

\end{deluxetable}

\begin{deluxetable}{lccccccc}
\tabletypesize{\scriptsize}
\tablecaption{Line EWs and upper limits for other optical lines predicted to trace the photoevaporative wind.\label{tab:optlines}}
\tablewidth{0pt}
\tablehead{
\colhead{Species} & \colhead{Wavelength} & \colhead{A$_{\rm ul}$\tablenotemark{a}}& \colhead{Terms}& \colhead{EW}&  \colhead{veiling}  &  \colhead{Luminosity\tablenotemark{a}}\\
\colhead{} & \colhead{(\AA)} & \colhead{(s$^{-1}$)} &\colhead{} &\colhead{(\AA)}& \colhead{} & \colhead{(L$_\sun$)}} 
\startdata
OI   & 5577.3387 & 1.26(0)&$^1$D-$^1$S&0.09 & 0.5 & 1.5(-6)\\
OI   & 6300.304  &  5.65(-3)&$^3$P-$^1$D&0.47 & 0.6 & 1.1(-5)\\
OI   & 6363.777  &1.82(-3)&$^3$P-$^1$D&0.15 & 0.4 & 3.2(-6)\\
OII & 7330.73 & 5.34(-2)&$^2$D-$^2$P&$<$0.04 & 0.5  & $<$9.1(-7)\\
MgI & 4571.0956 &2.54(2) &$^1$S-$^3$P&0.08& 0.5 & 6.4(-7)\\
NII  & 6583.460   &2.92(-3) &$^3$P-$^1$D&$<$0.007  & 0.4 & $<$1.5(-7)\\
SII   & 4068.600   &3.41(-1) &$^4$S-$^2$P&0.09 & 2 & 1.4(-6)\\
SII   & 4076.349   &1.34(-1) &$^4$S-$^2$P&$<$0.03      & 2 & $<$4.8(-7)\\
SII   & 6716.440  &4.66(-4)  &$^4$S-$^2$D&$<$0.01    & 0.5 & $<$2.3(-7)\\
SII   & 6730.810   &4.26(-4) &$^4$S-$^2$D&$<$0.008 & 0.6  & $<$1.9(-7)\\
\enddata
\tablenotetext{a}{The transition probability from NIST and the line luminosity are given as a(b) meaning a$\times$10$^b$.}
\tablecomments{Wavelengths are in air. EWs, veilings, and corresponding line luminosities are median values. Uncertainties on the measured EWs are about $\sim$10\%.
Because of night-to-night variations in the source brightness and in the veiling  the luminosities reported here have an uncertainty of $\sim$50\% (see Sect.~\ref{Sect:otherlines}).}
%
%
\end{deluxetable}

\bibliography{lit}

\end{document}